\documentclass[aps,prl,preprint,longbibliography,superscriptaddress]{revtex4}
\usepackage{CJK}
\usepackage{threeparttable}
\usepackage{indentfirst}
\usepackage{amsmath}
\usepackage{lipsum}
\usepackage{multirow}
\usepackage{booktabs}
\usepackage{graphicx,color}
\usepackage{mathrsfs}
\usepackage{graphicx}
\usepackage{subfigure}
\usepackage{amsmath}
\usepackage{mathtools}
\usepackage{braket}
\usepackage{color}

\newcommand{\ba}{\begin{eqnarray}}
\newcommand{\ea}{\end{eqnarray}}

\newcommand{\be}{\begin{equation}}
\newcommand{\ee}{\end{equation}}

\definecolor{pink}{rgb}{1,0.18,1.0}

\def\2dm{{2D Materials}}

\begin{document}

\title{Nonlinear harmonic spectra in bilayer van der Waals antiferromagnets CrX$_{3}$}

\author{Y. Q. Liu}
\affiliation{School of Materials and Energy, Lanzhou University, Lanzhou 730000, China}

\author{M. S. Si$^{*}$}
\affiliation{School of Materials and Energy, Lanzhou University, Lanzhou 730000, China}

\author{G. P. Zhang$^{\dagger}$}
\affiliation{Department of Physics, Indiana State University, Terre Haute, IN 47809, USA}

\date{\today}
\begin{abstract}
    Bilayer antiferromagnets CrX$_{3}$ (X $=$ Cl, Br, and I) are promising materials for spintronics and optoelectronics
    that are rooted in their peculiar electronic structures.
    However, their bands are often hybridized from the interlayer antiferromagnetic ordering, which are difficult to disentangle by traditional methods.
    In this work, we theoretically show that nonlinear harmonic spectra can differentiate subtle differences in their electronic states.
    In contrast to prior nonlinear optical studies which often use one or two photon energies,
    we systematically study the wavelength-dependent nonlinear harmonic spectra realized by hundreds of individual dynamical simulations under changed photon energies.
    Through turning on and off some excitation channels, we can pinpoint every dipole-allowed transition that largely contributes to the second and third harmonics.
    With the help of momentum matrix elements, highly entangled resonance peaks at a higher energy above the band edge
    can be assigned to specific transitions between the valence bands and three separate regions of conduction bands.
    Our findings demonstrate a feasible means to detect very complex electronic structures in an important family of two-dimensional antiferromagnets.
\end{abstract}

\maketitle
\section{I. Introduction}
Among two-dimensional van der Waals (vdW) magnetic materials discovered so far,
bilayers CrX$_{3}$ (X = Cl, Br, and I)
stand out as a key group of materials for spintronics and optoelectronics.
They exhibit remarkable electronic and optical properties \cite{CrCl-data,CrBr-data,CrI-data1,CrI-data2,CrI-data3} that are unparalleled by other materials.
The interlayer magnetic coupling between adjacent ferromagnetic monolayers of CrX$_{3}$ can be manipulated by applying
external fields \cite{electric2,electric3,pressure1,pressure2,pressure3} or adjusting the stacking patterns \cite{stacking1,stacking2}.
Different interlayer magnetic couplings in bilayers CrX$_{3}$ are vital to tunnel magnetoresistance.
More notably, bilayers CrX$_{3}$ provide an excellent platform for studying light-matter interactions such as magneto-optic Kerr effect \cite{MOKE},
magnetic circular dichroism \cite{RMCD}, polarized photoluminescence \cite{PL}, and second harmonic generation (SHG) \cite{shg1,shg2}.
Further progress is not possible without a detailed understanding of their electronic states.
Nonlinear optical techniques such as SHG, third or high harmonic generation (HHG) \cite{solid2,solid3,C60,reconstruct1,reconstruct2,reconstruct3}
and deliver attosecond laser pulses \cite{ferray1988}, are a method of choice.
Based on the three-step model proposed by Vampa \textit{et al}. \cite{three-step},
the valence and conduction bands during excitation are selected by the wavelength of a laser pulse.
They have shown that the interband contribution is more pronounced with a shorter wavelength \cite{three-step}.
In ZnO, the high harmonic spectra up to the 25th order are generated by the driving laser with wavelengths of 3250 nm,
where the maximum cutoff is determined by the maximum energy difference between the valence and conduction bands \cite{experiment-cutoff}.
Decreasing wavelength to 800 nm, Osika \textit{et al}. predict that the cutoff moves to the lower harmonics \cite{theory-cutoff}.
This wavelength dependence is confirmed in a model system \cite{inter-intra}.
In monolayer MoS$_{2}$, Liu \textit{et al}. report that the even- and odd-harmonics are extended to the 13th order under the wavelength of 4133 nm.
It is found that the even-order harmonic intensity is much smaller than that of the odd harmonic,
implying the different excited origins of these two kinds of harmonics \cite{solid3}.
This conclusion is also theoretically confirmed from the decreased wavelength \cite{MoS2}.
All the above studies reveal that the underlying physics of nonlinear optical response depends on the wavelength of the applied laser pulses.
In fact, both experimental and theoretical studies have already pointed out
that the wavelength dependence of harmonics is a key characteristics of solids,
which is different from the cases in atoms and small molecules.
In experiments, Sun et al. extracted SHG susceptibilities $|\chi^{(2)}_{xxx}|$ and $|\chi^{(2)}_{xyy}|$ of bilayer CrI$_{3}$
by scanning 17 excitation wavelengths in the range from 800 nm to 1040 nm \cite{shg1}.
At relatively low photon energies, the intensities of the resonance peaks are weak, but as the photon energy gradually increases, their intensities are strong.
However, due to a small number of photon energies used, some important resonance peaks may be missed, which is crucial for analyzing the electronic state of the material.
This motivates us to carry out a systematic study.

In this work, we demonstrate that nonlinear harmonics in a group of bilayers CrX$_{3}$ can pinpoint electronic states, state by state,
by scanning the photon energy between 0.3$-$1.4 eV in steps of 0.01 or 0.02 eV, rather than one or two photon energies often employed.
The first peaks of the 2nd harmonic spectra, resulting from the real electronic transitions between the valence and conduction bands,
are the onset of the optical band gaps for each CrX$_{3}$.
As the band gap gradually decreases with halogens down the periodic table, the peaks red shift.
Above the first peak, complex peaks are attributed to the resonance transitions.
We verify the reliability of the 2nd harmonic spectra by the continuous wave (cw) SHG.
For CrCl$_{3}$, the 2nd harmonic peaks, consisting of three eigenpeaks I, II, and III,
are associated with the conduction bands between 1.6$-$2.0 eV, called region 1 or $R_{1}$.
These three eigenpeaks are dominated by the transitions between the valence bands and the conduction bands in $R_{1}$.
The eigenpeaks result from the dipole-allowed transitions,
where the incident photon energy matches the energy difference between the valence and conduction bands,
and the corresponding momentum matrix elements are also large.
The same is true for the other two materials CrBr$_{3}$ and CrI$_{3}$.
In the 3rd harmonic spectra, fifteen transition channels are identified from the same strategy.
Our study demonstrates the potential power of nonlinear harmonic spectra
by mapping out the electronic states through a systematic scan of many incident photon energies, which complements the experimental studies \cite{shg1}.

The rest of the paper is arranged as follows.
In Sec. II, we show our theoretical methods.
Then, the results and discussions are given in Sec. III.
Section IV focuses on the third harmonic and the underlying physical picture.
Finally, we conclude our work in Sec. V.

\section{II. Theoretical Methods}

We employ the first-principles method combined with the Liouville equation to simulate the dynamical nonlinear optical response.
The electronic structures for bilayers CrX$_{3}$ are calculated in the framework of density functional theory (DFT), which is implemented in WIEN2k \cite{method1,method2,method3}.
We employ the generalized gradient approximation (GGA) for exchange-correlation potential in the form of Perdew-Burke-Ernzerhof (PBE) \cite{PBE}.
We self-consistently solve the Kohn-Sham equation \cite{KS1,KS2,KS3} as,
\begin{equation}\label{eq1}
\left[-\frac{\hbar^{2}}{2m_{e}}\nabla^{2}+V_{\rm eff}({\bf r})\right]\psi_{n{\bf k}}({\bf r})=
E_{n{\bf k}}\psi_{n{\bf k}}({\bf r}),
\end{equation}
where $\psi_{n\bf{k}}(\bf{r})$ is the Bloch wave function of band index $n$ at crystal
momentum $\bf{k}$, and $E_{n\bf{k}}$ denotes the corresponding eigenvalue.
The first and second terms of the Hamiltonian are the kinetic energy and the Kohn-Sham effective potential, respectively.
In our calculations, the plane wave expansion is determined by a cutoff parameter $R_{mt}K_{max}$ = 7.0.
Dense $k$-sample points with $32\times32\times1$ are used to ensure the convergence of the total energy.
All calculations are performed under the spin-orbit coupling (SOC),
which are realized through the second variational method.
The experimental lattice constants of 5.94, 6.30, and 6.87 {\AA} are used for CrCl$_{3}$, CrBr$_{3}$, and CrI$_{3}$, respectively \cite{CrBr-data,lattice1,lattice3}.
The AB-stacking order is taken for bilayers CrX$_{3}$.
The unit cell contains four Cr atoms and twelve X atoms.
The interlayer distances for CrCl$_{3}$, CrBr$_{3}$, and CrI$_{3}$ are 5.76, 6.05, and 6.82 {\AA}, respectively \cite{stacking2}.
To mimic the two-dimensional geometry, a vacuum layer of height 16 {\AA} is inserted between the bilayers along the $z$ direction.

To obtain HHG, we first construct the density matrix of the ground state as
$\rho_{0}=|\psi_{n{ \bf k}}({\bf r})\rangle\langle\psi_{n{\bf k}}({\bf r})|$.
Then, the dynamic density matrix can be obtained by numerically solving the time-dependent Liouville equation \cite{LV1,LV2,LV3}
\begin{equation}\label{eq2}
i \hbar\langle m{\bf k}|\partial \rho/\partial t|n{\bf k}\rangle=\langle m{\bf k}|[{\cal H},\rho]|n{\bf k}\rangle,
\end{equation}
where $\rho$ is the time-dependent density matrix, and a generic Hamiltonian is expressed as ${\cal H}={\cal H}_{0}+{\cal H}_{I}$.
${\cal H}_{0}$ is the system Hamiltonian.
${\cal H}_{I}$ is the interaction Hamiltonian between the laser field and the system:
${\cal H}_{I}=\frac{e}{m_{e}}\hat{{\bf P}}\cdot {\bf A}(t)$, where $\hat{{\bf P}}$
is the momentum operator, and ${\bf A}(t)$ represents the vector potential of the laser field.
The detailed vector potential used in this work is ${\bf A}(t)=A_{0}e^{-t^{2}/\tau^{2}}[$cos$(\omega t){\bf \hat{n}}]$
with ${\bf \hat{n}}$ being the unit vector of laser pulse polarization.
The vector potential amplitude is taken as $A_{0}$ = 0.03 Vfs/{\AA} and the duration $\tau$ is set to 60 fs.
The incident photon energies $\hbar\omega$ range from 0.3 to 1.4 eV at steps of 0.01 or 0.02 eV.
After obtaining the time-dependent density matrix $\rho$, we can calculate the expectation value of momentum operator \cite{mumentum} through
${\bf P}(t)=\Sigma_{{\bf k}}{\rm Tr}[\rho_{{\bf k}}(t)\hat{{\bf P}}_{{\bf k}}]$ with $\hat{{\bf P}}_{{\bf k}}=-i\hbar\nabla$.
Finally, we get the harmonic spectra by Fourier transforming ${\bf P}(t)$ into the frequency domain by
${\bf P}(\omega)=\int_{-\infty}^{\infty} {\bf P}(t)e^{i\omega t}{\cal W}(t)dt$
with $\omega$ being the harmonic frequency and ${\cal W}(t)$ the window function.

\section{III. Results and discussions}

\subsection{A. Static properties and cw SHG}
The crystal structure of bilayers CrX$_{3}$ is displayed in the bottom panel of Fig. 1(a).
Bilayers CrX$_{3}$ have S$_{6}$ point group.
There are six symmetry operations that can be generated by the inversion symmetry $\cal{I}$ and the threefold rotational symmetry $C_{3}$ with the trigonal axis as the $z$ axis.
It should be noted that $\cal{I}$ is broken from the interlayer antiferromagnetic (AFM) ordering with the magnetization of the monolayer along the $z$ axis, denoted as AFM-$z$.
Under AFM-$z$, the combined symmetry $\cal{IT}$ is preserved with $\cal{T}$ being the time reversal symmetry.
The band structure of CrCl$_{3}$ is given in Fig. 1(b).
All those bands are doubly degenerate, which is protected by $\mathcal{IT}$ \cite{Kramers}.
The valence bands near the Fermi level are mainly from the Cr-$d$ orbitals.
Below them, the bands mainly come from the Cl-$p$ orbitals.
Above the Fermi level, the conduction bands mainly come from the Cr-$d$ orbitals,
which are grouped into three separate regions $R_{1}$$-$$R_{3}$.

The nonlinear optical response has the potential to reveal the
physical properties of a material from the light-matter interaction.
The resonant enhancement is crucial to detect the band structure of a material,
and has the most important contribution to the susceptibilities \cite{P1,P2,P3}.
Based on the three-level model as schematically shown in the inset of Fig. 1(c), the second order susceptibility of cw SHG can be obtained as
\begin{equation}\label{eq1}
\begin{aligned}
\chi_{ijk}^{(2)}=&\frac{Ne^{3}}{2\epsilon_{0}\hbar^{2}}
\sum_{lmn}\rho_{ll}^{(0)} \bigg [ \frac{r_{ln}^{i}r_{nm}^{j}r_{ml}^{k}}{(\omega_{nl}-2\omega-i\eta)(\omega_{ml}-\omega-i\eta)}
+\frac{r_{ln}^{k}r_{nm}^{i}r_{ml}^{j}}{(\omega_{mn}-2\omega-i\eta)(\omega_{nl}+\omega+i\eta) }\\
+&\frac{r_{ln}^{j}r_{nm}^{i}r_{ml}^{k}}{(\omega_{nm}+2\omega+i\eta)(\omega_{ml}-\omega-i\eta) }
+\frac{r_{ln}^{k}r_{nm}^{j}r_{ml}^{i}}{(\omega_{ml}+2\omega+i\eta)(\omega_{nl}+\omega+i\eta) } \bigg] +(j \leftrightarrow k),
\end{aligned}
\end{equation}
where subscripts $i$, $j$, and $k$ of $\chi_{ijk}^{(2)}$ denote the Cartesian indices, $N$ is the atomic density,
$\rho_{ll}^{(0)}$ is the initial population for the state $|l \rangle$,
$\omega_{nl}=\omega_{n}-\omega_{l}$ with $\hbar\omega_{n}$ being the energy of band $n$, $\hbar\omega$ is the photon energy,
$r_{ln}$ is the matrix element of the position operator between electronic states $|l \rangle$ and $|n \rangle$, and
$\eta$ is the damping parameter \cite{damping1,damping2}.
The notation ($j \leftrightarrow k$) means an exchange of the two Cartesian directions.
The first term in Eq. (1) can be visualized by the double-sided Feynman diagram, as shown in the inset of Fig. 1(c).
The resonant enhancement can be understood from its denominator,
where two terms $(\omega_{nl}-2\omega-i\eta)$ and $(\omega_{ml}-\omega-i\eta)$ are involved.
When $2\hbar\omega=\hbar\omega_{nl}$ or $\hbar\omega=\hbar\omega_{ml}$, the cw SHG susceptibility has two typical peaks:
one at $\hbar\omega_{ml}$ for the transition $|m \rangle$ $\rightarrow$ $|l \rangle$
and the other at half of $\hbar\omega_{nl}$ for the transition $|n \rangle$ $\rightarrow$ $|l \rangle$.
Our peaks in cw SHG originate from these resonant contributions.
However, for the degenerate bands,
$\hbar\omega_{nl}$ and 2$\hbar\omega_{ml}$ are usually small.
This will lead to a divergence when the photon energy $\hbar\omega$ approaches zero \cite{MSHG}.
We are aware that the low-frequency divergence in the nonlinear optical response depends on the gauges used.
Although the velocity gauge can handle the low-frequency divergence,
it is problematic to truncate the summation of multiple bands \cite{exciton1,gauge1}.
On the other hand, the length gauge takes the $k$ derivative instead of position operator to account for the local effect of wavefunctions,
but to numerically compute it is challenging \cite{gauge2}.
Thus, our dynamical nonlinear response approach is a better way to overcome limitation, as discussed below.

\subsection{B. Dynamical response}
In the perturbation theory, the first-order density matrix $\rho^{(1)}$ is computed as
\be
i\hbar\dot{\rho}^{(1)}=[H_{0},\rho^{(1)}]+[H_{I},\rho^{(0)}],
\label{eq1}
\ee
where $H_{0}$ is the unperturbed Hamiltonian and $H_{I}$ is the interaction between the laser field and the system.
The second-order density matrix is given by
\be
i\hbar\dot{\rho}^{(2)}=[H_{0},\rho^{(2)}]+[H_{I},\rho^{(1)}].
\label{eq2}
\ee
In the interaction picture, $\rho^{(2)}$ is
\be
\rho^{(2)}(t)=\frac{1}{(i\hbar)^{2}}\int_{-\infty}^{t}dt_{1}\int_{-\infty}^{t_{1}}dt_{2}\big[H_{I}(t_{1}),[H_{I}(t_{2}),\rho_{0}]\big].
\label{eq3}
\ee
Because the system response is in the expectation value of the momentum operator through the trace Tr$(\rho\hat{\textbf{P}})$,
the second-order response depends on $\textbf{P}_{ij}$.
Although we numerically solve the Liouville equation exactly,
this dependence is useful for our purpose as illustrated in Fig. 1(a),
where the nonlinear response is linked to the product of multiple momentum transition matrix elements.
In addition, the nonlinear optical response based on the time-dependent density matrix has no limitation of the energy difference between electronic states.
This leads to a relatively accurate result near the zero photon frequency.
If the excitonic effects is considered, additional resonant peaks would appear in the low frequency range \cite{exciton2,exciton3},
which is beyond the scope of our current work.
A normal practice in HHG is to choose one or two photon energies and
then to Fourier transform the expectation value of momentum operator
${\bf P}(t)=\Sigma_{{\bf k}}{\rm Tr}[\rho_{{\bf k}}(t)\hat{{\bf P}}_{{\bf k}}]$ with $\hat{{\bf P}}_{{\bf k}}=-i\hbar\nabla$ to get the spectrum
as ${\bf P}(\omega)=\int_{-\infty}^{\infty} {\bf P}(t)e^{i\omega t}dt$.
But this has a deficiency.
Because of the energy-time uncertainty relation in the Fourier transform,
one cannot resolve harmonic peaks energetically down to 0.2 eV, except that one uses an extremely long pulse.
Higher harmonics need many more parameters, and
even within the perturbation theory their formulas become overly convoluted.
This requires a different strategy.

\subsection{C. Physics of even harmonics}
Nonmagnetic bilayers CrX$_{3}$ remain centrosymmetric as their monolayers.
If we consider the magnetic structures of Cr atoms, the inversion symmetry is broken from the interlayer antiferromagnetic order
as the top layer with spin-up magnetic orders are changed to the bottom layer with same spin-up magnetic orders
under the spatial inversion operator ($\vec{r}$$\rightarrow$$-\vec{r}$) and vice versa, as shown in the bottom panel of Fig. 1(a).
Meanwhile, the time reversal symmetry is also broken from this antiferromagnetic order.
According to the electric-dipole approximation, the even harmonics are allowed and also time nonreciprocal, usually denoted as c-type \cite{shg1}.
Our calculated HH spectra did include both the odd and even harmonics and confirm this symmetry analysis, as shown in Fig. S1(c) of the Supplemental Material (SM) \cite{SM}.
When the interlayer magnetic order is changed to be ferromagnetic, the spatial inversion symmetry is recovered
as the top layer with spin-up magnetic orders is identical to the bottom layer under $\vec{r}$$\rightarrow$$-\vec{r}$.
As a result, the even harmonics disappear, as shown in Fig. S1(d) \cite{SM}.
Thus, these unique magnetic structures in bilayers CrX$_{3}$ can be used to study the magnetization-mediated nonlinear harmonic spectra.

\subsection{D. Photon energy scan}
We scan the incident photon energies $\hbar\omega$ between 0.3 and 1.4 eV in step of 0.02 eV to obtain nonlinear harmonic spectra.
Figure 1(c) shows the 2nd harmonic spectra of bilayer CrCl$_{3}$.
When $\hbar\omega$ is below 0.9 eV, the 2nd harmonic signals are negligible.
We understand the reason behind. 2$\hbar\omega$ corresponds to the energy 1.8 eV, about 0.1 eV above the band gap of CrCl$_{3}$,
so only virtual transitions occur, which explains the negligible harmonic signal.
When the photon energy is above 0.9 eV, the real electron transitions touch the bands.
Thus, the first peak in this 2nd harmonic spectra can be regarded as an indicator of the band gap of a material.
This is also the case for CrBr$_{3}$ and CrI$_{3}$, as displayed in Figs. 1(d) and 1(e).

As the photon energy increases, additional six peaks 2$-$7 appear for CrCl$_{3}$, as shown in Fig. 1(c).
Four of them, peaks 2$-$4, and 6 are comparable to the first peak.
Peaks 5 and 7 have smaller amplitudes and peaks at 1.21 and 1.33 eV, respectively.
It should be noted that most of these characteristic peaks also appear in cw SHG, as discussed in SM \cite{SM} (see also Refs. \cite{X1,X2,X3,vasp1,vasp2} therein).
For CrBr$_{3}$, the 1st peak appears at $\hbar\omega=0.79$ eV, as shown in Fig. 1(d),
thus $2\hbar\omega=1.58$ eV is only 0.16 eV above the band gap of 1.42 eV for CrBr$_{3}$.
Seven peaks 2$-$8 appear after peak 1.
For CrI$_{3}$, more peaks appear, but have much stronger amplitudes, as shown in Fig. 1(e).

\subsection{E. Understanding the complex spectra}
The peaks in Figs. 1(c)$-$1(e) are very complex, but they offer an opportunity to demonstrate the power of harmonic generation.
In the following, we use an even finer step of 0.01 eV to scan the photon energy from 0.3 to 1.4 eV.
This requires 110 separate runs.
We focus on the 2nd harmonic spectra at $\Gamma$.
Figure 2(a) shows that the harmonic amplitude is small until the photon energy reaches 0.95 eV, where peaks I$-$III, called eigenpeak, appear.
Peak I has a small blueshift compared with the result in Fig. 1(c),
because the conduction band minimum and the valence band maximum of CrCl$_{3}$ are not at $\Gamma$.
Peak II appears at 1.13 eV, and its amplitude is large,
indicating an enhanced transition at this energy difference of 2.26 eV.
Peak III at 1.34 eV is very small.
The difference among these eigenpeak amplitudes implies the different transitions between the valence and conduction bands.
As seen above, we divide the conduction bands of CrCl$_{3}$ into three regions $R_{1}$$-$$R_{3}$,
so we can determine their separate effects on the harmonic amplitude.
We first exclude $R_{1}$ and find that peaks I and II disappear, and peak III decreases (the red line of Fig. 2(b)).
This means that peaks I$-$III are significantly affected by the transitions between the valence bands and conduction bands in $R_{1}$.
Excluding only $R_{2}$ hardly changes anything (the black line of Fig. 2(b)).
When only $R_{3}$ is excluded, these three peaks are also intact.
Therefore, eigenpeaks I$-$III in the 2nd harmonic spectra are mainly contributed by the transitions between the valence bands and conduction bands in $R_{1}$.

All the above discussions tell us that these eigenpeaks require resonant transitions,
where the incident photon energy must match the transition energy.
However, this is not the entire story.
Another criterion for these eigenpeaks is large momentum matrix elements between transition bands.
We calculate the momentum matrix elements for those bands that are indexed in 1$-$68 for bilayer CrCl$_{3}$, as shown in Fig. 2(c).
There are seven major regions with large momentum matrix elements,
which are labeled as A, B, C, D, E, F, and G, respectively.
We use lower case letters $a$, $b$, $c$, $d$, $e$, $f$, and $g$ for each transition in these regions, respectively.
Region A contains two main transitions $a_{1}$ and $a_{2}$, as shown in Fig. 2(d).
$a_{1}$ is the transition from the valence bands 35 and 36 to the conduction bands 41 and 42 in $R_{1}$,
where both the valence and conduction bands are doubly degenerate.
In Table I, we label it as 35,36$\rightarrow$41,42.
The transition energy of $a_{1}$ is about 2.02 eV, which matches the harmonic energy of peak I in Fig. 2(a).
The photon energy of peak I is about 0.95 eV.
Thus, $2\hbar\omega$ gives 1.90 eV and is nearly equal to the transition energy of $a_{1}$.
This means that peak I comes from the transition $a_{1}$.
The transition energy of $a_{2}$ is 2.16 eV, which is close to the 2nd harmonic energy of 2.26 eV for peak II.
Therefore, peak II comes from $a_{2}$.
There is only one transition $b$ in region B, and its transition energy is 2.66 eV.
It matches the 2nd harmonic energy of 2.68 eV for peak III, so peak III is dominated by $b$.
The $|\textbf{P}|$ for $a_{1}$, $a_{2}$, and $b$ are 0.03 a.u. (see Table I).
This shows that one can attribute each eigenpeak down to a specific transition, where the nonzero momentum matrix elements are essential.

\section{IV. Going beyond the second order and establishing a physical picture}
The above finding is also true for the 3rd harmonic spectra.
From the above discussion, a physical picture is emerging.
The harmonic response ${\cal R}$ is related to the momentum matrix products \cite{P2}
\be
{\cal R}^{(n)}(t)\propto\underbrace{\textbf{P}_{ij}\textbf{P}_{jk}\dots \textbf{P}_{lf}}_{n}e^{i(E_{f}-E_{i}-n\hbar\omega)t/\hbar}\cdots,
\label{eq4}
\ee
where $n$ is the order of interaction and the time-dependent exponent highlights the most important term only.
Obviously Eq. (5) is overly simplified \cite{P2},
but it captures the essential physics of the very complex processes during laser excitation as already seen in Eq. (4).
This response function replaces the susceptibility $\chi^{(n)}$ in the cw limit.
The finding is also true for the 3rd harmonic spectra.
Figure 3(a) shows the 3rd harmonic spectra at $\Gamma$ for CrCl$_{3}$,
where the photon energy is scanned from 0.5 to 1.4 eV in step of 0.01 eV.
The first peak is at 1.02 eV, followed by six additional peaks 2$-$7.
Peaks 1$-$3, 6, and 7 have comparable amplitudes, while peaks 4 and 5 are relatively small.
Figure 3(b) presents two sets of data: one excludes $R_{1}$ and the other excludes $R_{2}$.
When only $R_{1}$ is excluded (the red line), the original peaks 1, 2, and 3 disappear.
Peaks 6 and 7 are largely reduced, while peaks 4 and 5 are almost unchanged.
This indicates that peaks 1$-$3, 6, and 7 are dominated by the transitions between the valence bands and conduction bands in $R_{1}$.
When we only exclude $R_{2}$ (the black line), peaks 1$-$3, 6, and 7 remain almost unchanged,
but peaks 4 and 5 are sharply reduced.
This means that those transitions are closely related to peaks 4 and 5.
When we only exclude $R_{3}$,
seven peaks are unchanged, indicating that these peaks are dominated by $R_{1}$ and $R_{2}$.

Recall Fig. 2(c), where we have five regions C$-$G with large momentum matrix elements $|\textbf{P}|$ between the valence bands and conduction bands in $R_{1}$ and $R_{2}$.
Figure 3(c) shows the energy diagram and five regions from C to G, each of which has multiple transitions.
Region C contains four transitions: $c_{1}$, $c_{2}$, $c_{3}$, and $c_{4}$, as shown in Fig. 3(c) and Table I.
The transition energy of $c_{1}$ is about 3.07 eV, which matches the 3rd harmonic energy of 3.06 eV for peak 1 in Fig. 3(a).
This shows that peak 1 mainly comes from the transition $c_{1}$.
The transition energy of $c_{2}$ is only 0.01 eV larger than that of $c_{1}$, so it also contributes to peak 1.
Comparing with $c_{1}$ and $c_{2}$, $c_{3}$ and $c_{4}$ have slightly larger transition energies of 3.14 and 3.15 eV,
which generate peak 2 with $3\hbar\omega=3.15$ eV.
The absolute value of $|\textbf{P}|$ for $c_{1}$ and $c_{2}$ is about 0.18 and 0.15 a.u. (see Table \uppercase\expandafter{\romannumeral1}).
The other two transitions $c_{3}$ and $c_{4}$ have almost equal $|\textbf{P}|$ with the values of 0.13 and 0.15 a.u., respectively,
which explains their comparable amplitudes of peaks 1 and 2, as shown in Fig. 3(a).
Region D contains two transitions $d_{1}$ and $d_{2}$, as shown in Fig. 3(c).
The corresponding transition energies are 3.20 eV and 3.21 eV,
which contribute to peak 2.
The transitions $d_{1}$ and $d_{2}$ have a smaller $|\textbf{P}|$ of 0.11 a.u..
Region E has only one transition $e$, and the corresponding transition energy is 3.26 eV.
It matches the harmonic energy of peak 3, where $3\hbar\omega=3.36$ eV.
The $|\textbf{P}|$ of this transition is 0.09 a.u., which is smaller than those of peaks 1 and 2.
This leads to the amplitude of peak 3 smaller than those of peaks 1 and 2.
For region F, there are four transitions $f_{1}$$-$$f_{4}$ with the transition energies of 3.88, 3.89, 3.98, and 3.99 eV, respectively.
These transition energies are close to the harmonic energies of peaks 6 and 7 with $3\hbar\omega=3.90$ and 3.99 eV,
indicating that these two peaks are mainly from the transitions $f_{1}$$-$$f_{4}$ (see Fig. 3(c)).
The corresponding $|\textbf{P}|$ are 0.11, 0.11, 0.12, and 0.11 a.u., respectively.
It is found that the $|\textbf{P}|$ of peaks 6 and 7 is almost equal, so the amplitudes of these two peaks are also close, as shown in Fig. 3(a).

As shown in Fig. 2(c), we can see that only one region G related to $R_{2}$ has large momentum matrix elements.
It mainly contains four transitions $g_{1}$, $g_{2}$, $g_{3}$, and $g_{4}$, as shown in Fig. 3(c) and Table \uppercase\expandafter{\romannumeral1}.
Their transition energies are 3.46, 3.53, 3.56, and 3.65 eV, respectively.
The first three transition energies almost match the harmonic energy of peak 4 with $3\hbar\omega=3.45$ eV.
This indicates that these three transitions contribute to peak 4.
This is the reason why peak 4 is so broad.
However, only one transition $g_{4}$ contributes to peak 5 with $3\hbar\omega=3.69$ eV,
giving rise to a narrow peak in comparison with peak 4.
In addition, the momentum matrix elements $|\textbf{P}|$ of $g_{1}$$-$$g_{3}$ are 0.23, 0.20, and 0.13 a.u., respectively.
By contrast, $g_{4}$ has a slightly smaller $|\textbf{P}|$ of 0.11 a.u..
As a result, the amplitude of peak 5 is smaller than that of peak 4.

It is found that the amplitudes of eigenpeaks for the 3rd harmonic spectra are larger than those for the 2nd harmonic spectra.
The appearance of the strong eigenpeaks needs large $|\textbf{P}|$.
In fact, $|\textbf{P}|$ in regions C$-$G did have much larger values than those in regions A and B, as shown in Fig. 2(c).
The underlying physics comes from the fact that $|\textbf{P}|$ contributing to the 2nd harmonic spectra are related to the
valence bands 29$-$40 and the conduction bands in $R_{1}$, which are mainly dominated by the same Cr-$d$ orbitals, as shown in Fig. 1(b).
However, $|\textbf{P}|$ for the 3rd harmonic spectra involves a larger energy difference.
The corresponding valence bands are much lower, which are mainly from the Cl-$p$ orbitals
and are denoted as the bands 1$-$28 in Fig. 1(b).
Compared to $|\textbf{P}|$ related to the 2nd harmonic spectra,
$|\textbf{P}|$ for the 3rd harmonic spectra would be much larger as the transitions belong to different orbitals.
Our calculated momentum matrix elements confirm this conclusion (see Table I for more details).

\section{V. Conclusions}
We have demonstrated that the nonlinear harmonic spectra in a group of bilayers CrX$_{3}$ can pinpoint specific transitions between valence and conduction bands.
We scan over 100 photon energies from 0.3$-$1.4 eV in fine steps.
We map out the momentum transition matrix elements and calculate the corresponding contributions to the harmonic spectra.
By including and excluding some transitions, we can directly correlate each peak in both the 2nd and 3rd harmonic spectra to a specific transition,
which has not been possible previously because often time only one or two photon energies are used.
The first resonance peak in the 2nd harmonic spectra serves as the onset of the band gap of a material.
Complex peaks can now be attributed to the intrinsic electronic properties.
All the results are further confirmed by the 3rd harmonic spectra.
Our findings unleash the power of nonlinear harmonic spectra to study the vdW antiferromagnets,
and are likely to motivate future experimental studies beyond CrI$_{3}$ \cite{shg1}.

\section*{ACKNOWLEDGMENTS}
This work was supported by the National Science Foundation of China under Grant No. 11874189.
GPZ was supported by the U.S. Department of Energy under Contract No. DE-FG02-06ER46304.
The research used resources of the National Energy
Research Scientific Computing Center, which is supported by
the Office of Science of the U.S. Department of Energy under
Contract No. DE-AC02-05CH11231.

$^{*}$sims@lzu.edu.cn

$^{\dagger}$guo-ping.zhang@outlook.com

\clearpage

\begin{table}[bh]
\centering
\caption{Transitions in regions A$-$G in Fig. 2(c), with large momentum matrix elements $|\textbf{P}|$.
The corresponding transition energies match the harmonic energies of the eigenpeaks,
where I$-$III denote the peaks in the 2nd harmonic spectra and 1$-$7 denote those eigenpeaks in the 3rd harmonic spectra.
$\Delta$$E$ and $\hbar\omega$ represent the transition and photon energies, respectively, in units of eV.
The momentum matrix elements are in units of atomic unit (a.u.).
}
\begin{tabular}{cccccccccccccccccccccccccc}\hline\hline
Region                  &&Transition                &&$|\textbf{P}|$ &&$\Delta$$E$                  && Peak                    && $\hbar\omega$ (2$\hbar\omega$/3$\hbar\omega$) \\ \hline
\multirow{2}{0.2cm}{A}  &&35,36$\rightarrow$41,42   &&0.03    &&\multirow{1}{1.5cm}{2.02 ($a_{1}$)} &&\multirow{1}{0.2cm}{I}   &&\multirow{1}{1.8cm}{0.95 (1.90)}                \\
                        &&31,32$\rightarrow$43,44   &&0.03    &&\multirow{1}{1.5cm}{2.16 ($a_{2}$)} &&\multirow{1}{0.2cm}{II}   &&\multirow{1}{1.8cm}{1.13 (2.26)}                 \\\hline
\multirow{1}{0.2cm}{B}  &&29,30$\rightarrow$47,48   &&0.03    &&\multirow{1}{1.5cm}{2.66 ($b$)}       &&\multirow{1}{0.2cm}{III}   &&\multirow{1}{1.8cm}{1.34 (2.68)}                  \\\hline
\multirow{4}{0.2cm}{C}  &&23,24$\rightarrow$45,46   &&0.18    &&\multirow{1}{1.5cm}{3.07 ($c_{1}$)} &&\multirow{1}{0.2cm}{1}   &&\multirow{2}{1.8cm}{1.02 (3.06)}                  \\
                        &&23,24$\rightarrow$47,48   &&0.15    &&\multirow{1}{1.5cm}{3.08 ($c_{2}$)} &&\multirow{1}{0.2cm}{1}   &&                                                   \\
                        &&21,22$\rightarrow$45,46   &&0.13    &&\multirow{1}{1.5cm}{3.14 ($c_{3}$)} &&\multirow{1}{0.2cm}{2}   &&\multirow{2}{1.8cm}{1.05 (3.15)}                    \\
                        &&21,22$\rightarrow$47,48   &&0.15    &&\multirow{1}{1.5cm}{3.15 ($c_{4}$)} &&\multirow{1}{0.2cm}{2}   &&                                                     \\\hline
\multirow{2}{0.2cm}{D}  &&11,12$\rightarrow$41,42   &&0.11    &&\multirow{1}{1.5cm}{3.20 ($d_{1}$)} &&\multirow{1}{0.2cm}{2}   &&\multirow{2}{1.8cm}{1.05 (3.15)}                      \\
                        &&13,14$\rightarrow$43,44   &&0.11    &&\multirow{1}{1.5cm}{3.21 ($d_{2}$)} &&\multirow{1}{0.2cm}{2}   &&                                                       \\\hline
\multirow{1}{0.2cm}{E}  &&15,16$\rightarrow$48,47   &&0.09    &&\multirow{1}{1.5cm}{3.26 ($e$)}       &&\multirow{1}{0.2cm}{3}   &&\multirow{1}{1.8cm}{1.12 (3.36)}                        \\\hline
\multirow{4}{0.2cm}{F}  &&7,8$\rightarrow$45,46     &&0.11    &&\multirow{1}{1.5cm}{3.88 ($f_{1}$)} &&\multirow{1}{0.2cm}{6}   &&\multirow{2}{1.8cm}{1.30 (3.90)}                         \\
                        &&7,8$\rightarrow$47,48     &&0.11    &&\multirow{1}{1.5cm}{3.89 ($f_{2}$)} &&\multirow{1}{0.2cm}{6}                      &&                                       \\
                        &&5,6$\rightarrow$45,46     &&0.12    &&\multirow{1}{1.5cm}{3.98 ($f_{3}$)} &&\multirow{1}{0.2cm}{7}   &&\multirow{2}{1.8cm}{1.33 (3.99)}                           \\
                        &&5,6$\rightarrow$47,48     &&0.11    &&\multirow{1}{1.5cm}{3.99 ($f_{4}$)} &&\multirow{1}{0.2cm}{7}                  &&                                             \\\hline
\multirow{4}{0.2cm}{G}  &&27,28$\rightarrow$51,52   &&0.23    &&\multirow{1}{1.5cm}{3.46 ($g_{1}$)} &&\multirow{1}{0.2cm}{4}   &&\multirow{3}{1.8cm}{1.15 (3.45)}                             \\
                        &&25,26$\rightarrow$49,50   &&0.20    &&\multirow{1}{1.5cm}{3.53 ($g_{2}$)} &&\multirow{1}{0.2cm}{4}   &&                                                              \\
                        &&27,28$\rightarrow$55,56   &&0.13    &&\multirow{1}{1.5cm}{3.56 ($g_{3}$)} &&\multirow{1}{0.2cm}{4}   &&                                                               \\
                        &&25,26$\rightarrow$55,56   &&0.11    &&\multirow{1}{1.5cm}{3.65 ($g_{4}$)} &&\multirow{1}{0.2cm}{5}   &&\multirow{1}{1.8cm}{1.23 (3.69)}                                \\
\hline\hline
\end{tabular}\label{tab1}
\end{table}
\clearpage

\begin{figure}
\centering
\includegraphics[width=.85\textwidth]{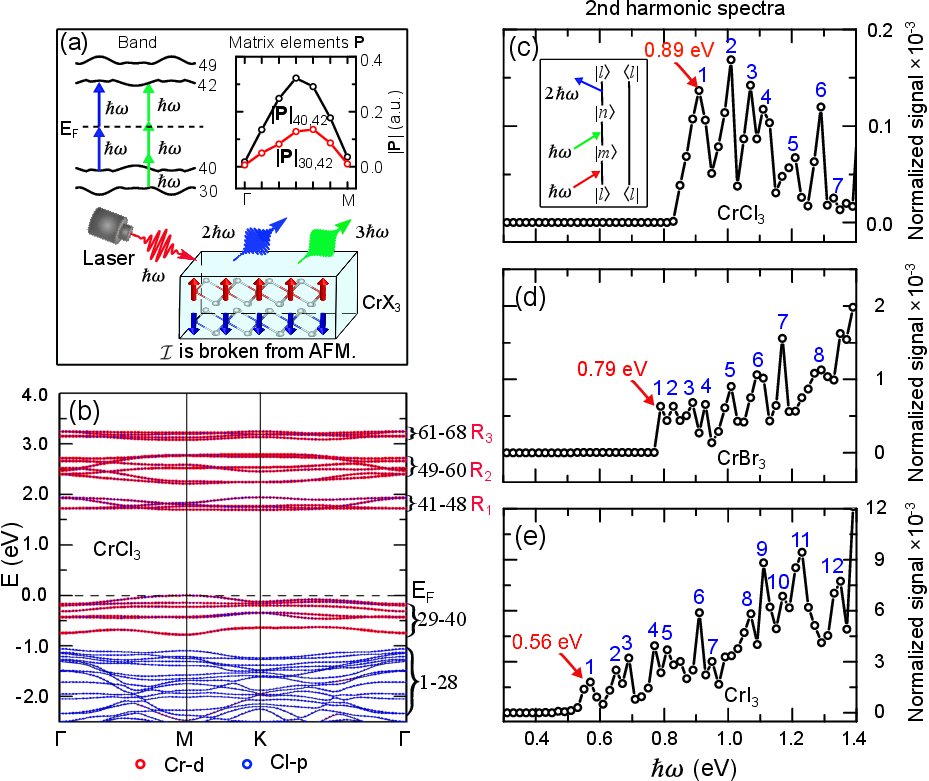}
\caption{(a) A schematic diagram of nonlinear optical responses in bilayer CrX$_{3}$, which are related
to the band structure and the momentum matrix elements \textbf{P}.
$|\textbf{P}|$$_{40,42}$ between the bands 40 and 42 is larger than $|\textbf{P}|$$_{30,42}$ along $\Gamma$$-$$M$ in bilayer CrCl$_{3}$.
(b) Orbital-resolved band structure for bilayer CrCl$_{3}$,
where the red and blue circles represent the Cr-$d$ and Cl-$p$ orbitals, respectively.
The band indices 1$-$68 are marked on the right side of this figure.
The conduction bands are divided into three individual regions $R_{1}$$-$$R_{3}$.
(c) The 2nd harmonic spectra as a function of photon energy for bilayer CrCl$_{3}$.
The blue numbers near the peaks represent the peak indices and the red number is the photon energy of the first peak.
The inset shows the conventional double-sided Feynman diagram corresponding to the first term in Eq. (1).
The Feynman diagram has two lines of propagation, one for the $|l \rangle$ side of $\rho^{(0)}_{ll}$ and the other for the $\langle l|$ side.
The red, green, and blue arrows correspond to the input photon energies $\hbar\omega$ and $\hbar\omega$ and the output photon energies $2\hbar\omega$, respectively.
The energy conservation relation $\hbar\omega$ + $\hbar\omega$ = $2\hbar\omega$ is satisfied.
(d) and (e) The 2nd harmonic spectra as a function of photon energy for bilayers CrBr$_{3}$ and CrI$_{3}$, respectively.
}
\label{fig1}
\end{figure}

\clearpage

\begin{figure}
\centering
\includegraphics[width=.9\textwidth]{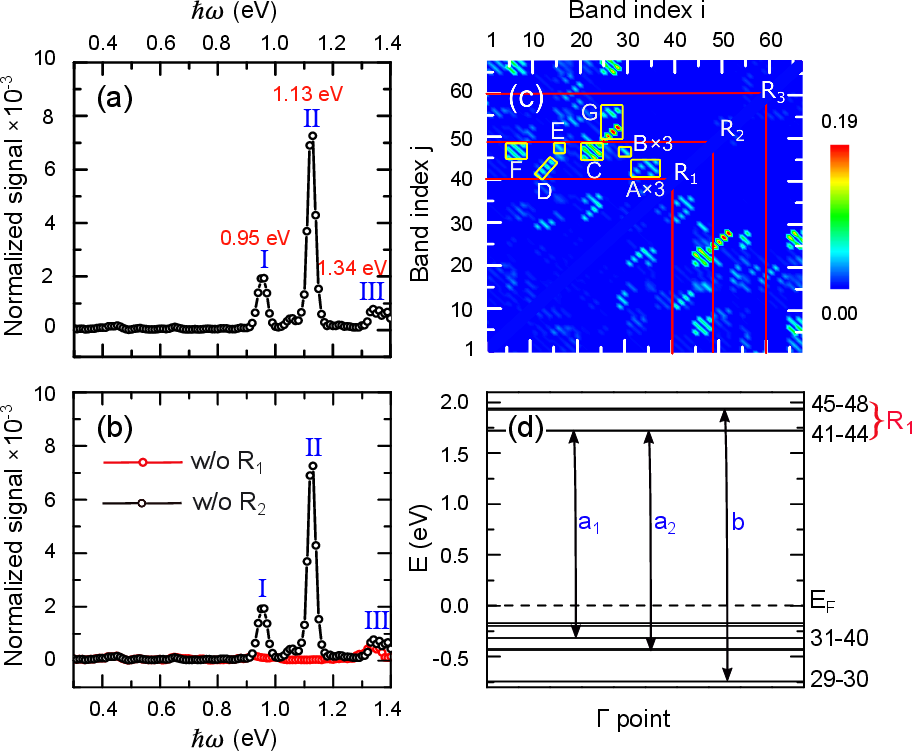}
\caption{(a) The 2nd harmonic spectra as a function of photon energy at $\Gamma$ for bilayer CrCl$_{3}$.
The blue and red numbers near each peak represent the peak indices and the corresponding photon energies, respectively.
(b) The 2nd harmonic spectra as a function of photon energy excluding $R_{1}$ (red line) and $R_{2}$ (black line) in the conduction bands.
(c)  The absolute value of the momentum matrix elements $|\textbf{P}|$ for bands 1$-$68 in bilayer CrCl$_{3}$.
(d) The energy diagram at $\Gamma$ for bilayer CrCl$_{3}$,
where the arrows denote transitions with large $\textbf{P}$,
and the letters on the arrow label the different transitions.}
\label{fig2}
\end{figure}

\clearpage

\begin{figure}
\centering
\includegraphics[width=.9\textwidth]{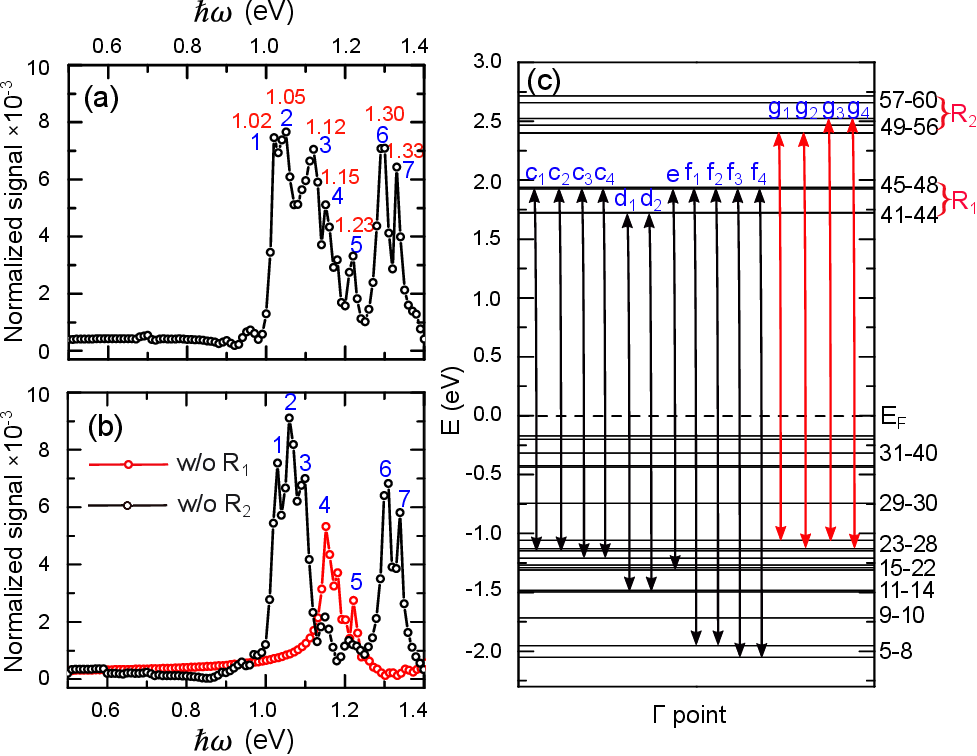}
\caption{
(a) The 3rd harmonic spectra as a function of photon energy $\hbar\omega$ at $\Gamma$.
The blue and red numbers near each peak represent the peak indices and the photon energies, respectively.
(b) The 3rd harmonic spectra as a function of photon energy excluding $R_{1}$ (red line) and $R_{2}$ (black line) in the conduction bands.
(c) The energy diagram at $\Gamma$ for bilayer CrCl$_{3}$,
where the arrows denote transitions with large $\textbf{P}$,
and the letters on the arrow label different transitions.
}
\label{fig3}
\end{figure}
\end{document}

\title{Supplemental Material for ``Nonlinear harmonic spectra in bilayer van der Waals antiferromagnets CrX$_{3}$"}

\author{Y. Q. Liu}
\affiliation{School of Materials and Energy, Lanzhou University, Lanzhou 730000, China}

\author{M. S. Si$^{*}$}
\affiliation{School of Materials and Energy, Lanzhou University, Lanzhou 730000, China}

\author{G. P. Zhang$^{\dagger}$}
\affiliation{Department of Physics, Indiana State University, Terre Haute, IN 47809, USA}

\date{\today}

\maketitle

This Supplemental Material includes five sections.
In Sec. \uppercase\expandafter{\romannumeral1}, we briefly introduce theoretical method used in our work.
In Sec. \uppercase\expandafter{\romannumeral2}, the electronic structures and high harmonic generation (HHG) for bilayers CrX$_{3}$ are given.
In Sec. \uppercase\expandafter{\romannumeral3}, we demonstrate the results of cw second-harmonic generation (SHG) and the 3rd harmonic spectra for CrX$_{3}$.
In Sec. \uppercase\expandafter{\romannumeral4}, we discuss the underlying physics of eigenpeaks for the nonlinear harmonic spectra of CrBr$_{3}$.
Finally, the underlying physics of eigenpeaks in the nonlinear harmonic spectra for CrI$_{3}$ is given in Sec. \uppercase\expandafter{\romannumeral5}.

\section{I. Theoretical Methods}
To calculate the cw SHG of semiconductor CrX$_{3}$, we used the length-gauge formulation of SHG susceptibility developed by Sipe \cite{X1,X2,X3}.
The second nonlinear polarization is given as
$P^{a}(2\omega)=\chi^{abc}(2\omega;\omega,\omega)\varepsilon^{b}(\omega)\varepsilon^{c}(\omega)$,
where $\chi^{abc}$ is the SHG susceptibility
and $\varepsilon^{b}$ denotes the $b$ component of the optical electric field.
$\chi^{abc}$ mainly contains three parts: the interband transitions $\chi^{abc}_{inter}$,
the intraband transitions $\chi^{abc}_{intra}$, and the modulation of interband terms by intraband terms $\chi^{abc}_{mod}$, where $\chi^{abc}_{mod}$ is very small.
The specific expressions are
\begin{align}
\chi^{abc}_{inter}(2\omega;\omega,\omega)=&\frac{e^{3}}{\hbar^{2}}\int\frac{d\textbf{k}}{4\pi^{3}}\sum_{nml}
\left\{\frac{2f_{nm}}{\omega_{mn}(\textbf{k})-2\omega}+\frac{f_{ml}}{\omega_{ml}(\textbf{k})-\omega}+\frac{f_{ln}}{\omega_{ln}(\textbf{k})-\omega}\right\}\notag\\
&\times\frac{r_{nm}^{a}(\textbf{k})\left\{r_{ml}^{b}(\textbf{k})r_{ln}^{c}(\textbf{k})\right\}}
{\omega_{ln}(\textbf{k})-\omega_{ml}(\textbf{k})}, \tag{3}\\
\chi^{abc}_{intra}(2\omega;\omega,\omega)=&\frac{e^{3}}{\hbar^{2}}\int\frac{d\textbf{k}}{4\pi^{3}}\sum_{nml}\omega_{mn}(\textbf{k})
r_{nm}^{a}(\textbf{k})\left\{r_{ml}^{b}(\textbf{k})r_{ln}^{c}(\textbf{k})\right\}\notag\\
&\times \bigg\{\frac{f_{nl}}{\omega_{ln}^{2}(\textbf{k})\left(\omega_{ln}(\textbf{k})-\omega\right)}
-\frac{f_{lm}}{\omega_{ml}^{2}(\textbf{k})\left(\omega_{ml}(\textbf{k})-\omega\right)}\bigg\} \notag\\
&-\frac{8ie^{3}}{\hbar^{2}}\int\frac{d\textbf{k}}{4\pi^{3}}\sum_{nm}
\frac{f_{nm}r_{nm}^{a}(\textbf{k})}{\omega_{mn}^{2}(\textbf{k})(\omega_{mn}(\textbf{k})-2\omega)}\left\{\Delta_{mn}^{b}(\textbf{k})r_{mn}^{c}(\textbf{k})\right\} \notag\\
&-\frac{2e^{3}}{\hbar^{2}}\int\frac{d\textbf{k}}{4\pi^{3}}\sum_{nml}\frac{f_{nm}r_{nm}^{a}
\left(\omega_{ln}(\textbf{k})-\omega_{ml}(\textbf{k})\right)\left\{r_{ml}^{b}(\textbf{k})r_{ln}^{c}(\textbf{k})\right\}}{\omega_{mn}^{2}(\textbf{k})(\omega_{mn}(\textbf{k})-2\omega)}, \tag{4}\\
\frac{i}{2\omega}\chi^{abc}_{mod}(2\omega;\omega,\omega)=&\frac{ie^{3}}{2\hbar^{2}}\int\frac{d\textbf{k}}{4\pi^{3}}\sum_{nml}
\frac{\omega_{nl}(\textbf{k})r_{lm}^{a}(\textbf{k})\left\{r_{mn}^{b}(\textbf{k})r_{nl}^{c}(\textbf{k})\right\}f_{nm}}{\omega_{mn}^{2}(\textbf{k})\left(\omega_{mn}(\textbf{k})-\omega\right)}\notag\\
&-\frac{ie^{3}}{2\hbar^{2}}\int\frac{d\textbf{k}}{4\pi^{3}}\sum_{nml}\frac{\omega_{lm}(\textbf{k})r_{nl}^{a}(\textbf{k})\left\{r_{lm}^{b}(\textbf{k})r_{mn}^{c}(\textbf{k})\right\}f_{nm}}{\omega_{mn}^{2}(\textbf{k})\left(\omega_{mn}(\textbf{k})-\omega\right)}\notag\\
&+\frac{ie^{3}}{2\hbar^{2}}\int\frac{d\textbf{k}}{4\pi^{3}}\sum_{nm}\frac{f_{nm}\Delta_{nm}^{a}(\textbf{k})\left\{r_{mn}^{b}(\textbf{k})r_{nm}^{c}(\textbf{k})\right\}}
{\omega_{mn}^{2}(\textbf{k})\left(\omega_{mn}(\textbf{k})-\omega\right)}, \tag{5}
\end{align}
where $\omega_{mn}(\textbf{k})=\omega_{m}(\textbf{k})-\omega_{n}(\textbf{k})$, and $\hbar\omega_{n}$ represents the energy of band $n$.
Note that $r_{mn}(\textbf{k})$ is the matrix element of position operator obtained by the first-principles calculations,
as implemented in the Vienna ab initio simulation package (VASP) \cite{vasp1,vasp2}.
GGA in the PBE implementation is applied as the exchange correlation functional.
To accurately obtain $\chi^{abc}$, a large number of $k$ points and energy bands are required.
The dense $k$-point mesh of $32\times32\times1$ is used, and the number of energy bands is set to 132.

\section{\uppercase\expandafter{\romannumeral2}. The electronic structure and high harmonic generation from bilayers CrX$_{3}$ }
For bilayer CrBr$_{3}$, the band gap is about 1.42 eV, as shown in Fig. S1(a), which is 0.27 eV smaller than that of CrCl$_{3}$.
The valence bands in the energy range of -0.5 to 0.0 eV and
the conduction bands of 1.4 to 3.5 eV are both occupied by Cr-$d$ orbitals,
while the valence bands of -3.0 to -0.5 eV are occupied by Br-$p$ orbitals.
Bilayer CrI$_{3}$ is a direct-gap semiconductor with a band gap about 0.87 eV, as shown in Fig. S1(b).
The band gap of CrI$_{3}$ is smaller than those of the other two materials.
The valence bands of CrI$_{3}$ are mainly occupied by I-$p$ orbitals, while the conduction bands are mainly occupied by Cr-$d$ orbitals.
The hybridization between the I-$p$ and Cr-$d$ orbitals increases, so these two orbitals gradually approach the Fermi level,
resulting in a decrease in the band gap.
It is found that the conduction bands of bilayers CrBr$_{3}$ and CrI$_{3}$ can also be divided into three individual regions $R_{1}$$-$$R_{3}$,
which is similar to bilayer CrCl$_{3}$ (see the main text).
These unique electronic structures dominate the nonlinear optical response.

\begin{figure}[th]
\centering
\includegraphics[width=.65\textwidth]{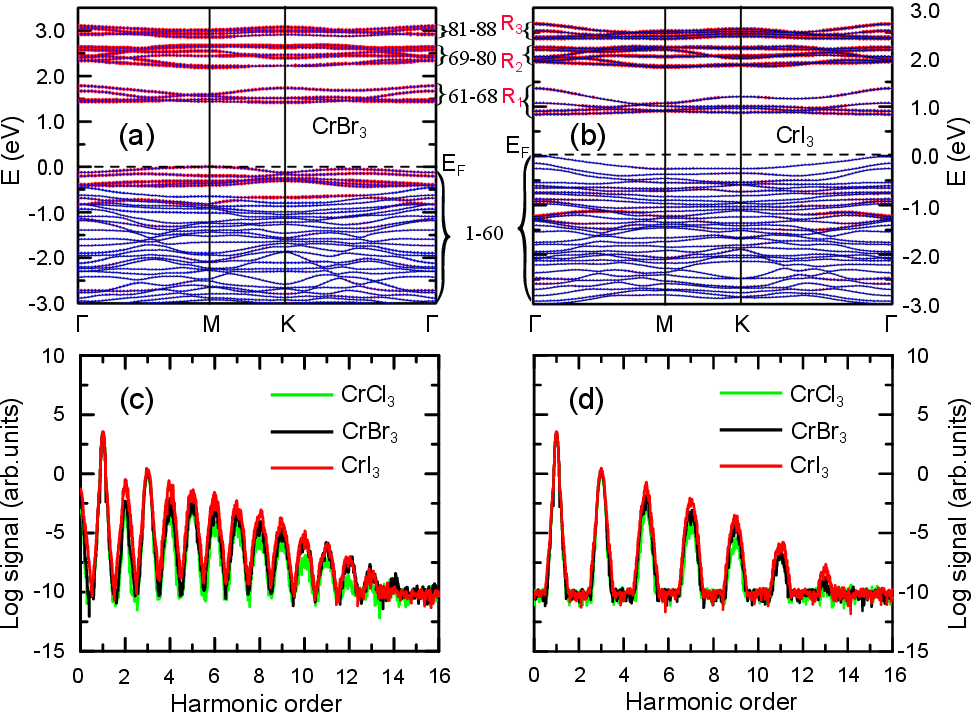}
\caption{ Orbital-resolved band structures for bilayers (a) CrBr$_{3}$ and (b) CrI$_{3}$,
where red circles represent the Cr-$d$ orbitals and blue circles represent Br-$p$ and I-$p$ orbitals.
The band indices 1$-$88 are marked in figures (a) and (b).
The conduction bands can be divided into three individual regions $R_{1}$$-$$R_{3}$.
$R_{1}$ contains the bands 61$-$68, $R_{2}$ the bands 69$-$80,
and $R_{3}$ the bands 81$-$88.
The dashed line denotes the Fermi level.
HHGs from bilayers CrCl$_{3}$ (green line), CrBr$_{3}$ (black line), and CrI$_{3}$ (red line)
with (c) AFM and (d) ferromagnetic coupling. The laser polarization along the $x$ axis.
The photon energy is $\hbar\omega$ = 0.5 eV, with the duration $\tau$ = 60 fs, and the vector potential amplitude $A_{0}$ = 0.03 Vfs/{\AA}.
}
\label{S1}
\end{figure}

\begin{figure}
\centering
\includegraphics[width=.4\textwidth]{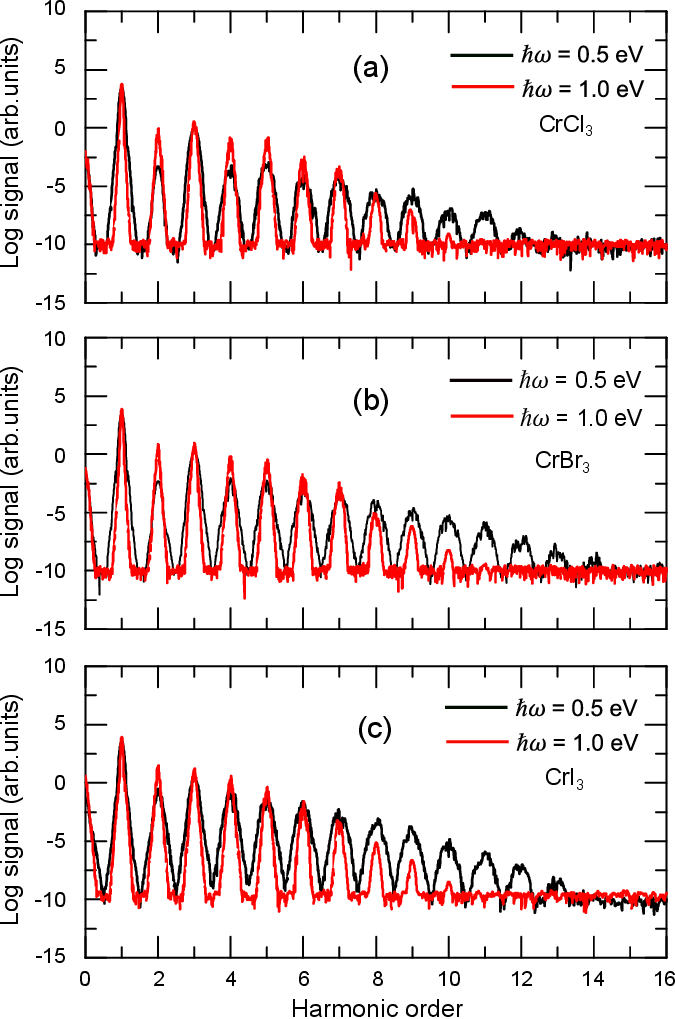}
\caption{
HHG signal from bilayers (a) CrCl$_{3}$, (b) CrBr$_{3}$ and (c) CrI$_{3}$,
where the black and red lines denote $\hbar\omega$ = 0.5 and 1.0~eV, respectively.
The laser polarization along the $x$ axis.
The photon energy is $\hbar\omega$ = 0.5 eV, with the duration $\tau$ = 60 fs,
and the vector potential amplitude $A_{0}$ = 0.03 Vfs/{\AA}.
HHGs are plotted on the logarithmic scale.}
\label{S1}
\end{figure}

Under the linearly polarized laser along the $x$ axis, HHGs for these three materials are given in Fig. S1(c),
where both even and odd harmonics appear.
The appearance of even harmonics originates from the broken inversion symmetry.
Here, we only focus on the component parallel to the laser field.
We can see that the generated harmonics reach as high as the 14th order.
For the same laser's parameters, HHGs are different for these three materials, in particular for those even orders.
For instance, CrI$_{3}$ has a strongest amplitude for the 2nd harmonics, while
that of CrCl$_{3}$ is smallest, which are inversely proportional to their band gaps.
However, this amplitude difference is very small for those odd harmonics.
For ferromagnetic CrX$_{3}$, only the odd harmonics appear, as shown in Fig. S1(d).
This is because the existence of $\cal{I}$ leads to the disappearance of even harmonics.

In addition, it is found that photon energy can further tune the harmonic amplitudes for these three materials.
In the case of CrCl$_{3}$, the photon energy of 1.0 eV gives a stronger harmonic amplitude than that of 0.5 eV for
the low order harmonics, but smaller for those high order harmonics, as shown in Fig. S2(a).
This is also the case for other two materials, as shown in Figs. S2(b) and S2(c).
Therefore, the harmonic signal largely depends on the photon energy.
This needs to scan the photon energy to obtain nonlinear harmonic spectra.

\section{\uppercase\expandafter{\romannumeral3}. The cw SHG and the 3rd harmonic spectra of CrX$_{3}$}

To confirm the reliability of the 2nd harmonic spectra,
we have calculated the cw SHG in bilayers CrX$_{3}$.
For AFM-$z$, bilayers CrX$_{3}$ only have the threefold rotational symmetry $C_{3}$.
As a result, there exist two independent nonvanishing elements:
$\chi_{xxx}^{(2)}=-\chi_{xyy}^{(2)}=-\chi_{yxy}^{(2)}=-\chi_{yyx}^{(2)}$, and
$\chi_{xxy}^{(2)}=\chi_{xyx}^{(2)}=\chi_{yxx}^{(2)}=-\chi_{yyy}^{(2)}$.
The SHG susceptibilities satisfy the intrinsic permutation symmetry $\chi_{abc}^{(2)}=\chi_{acb}^{(2)}$.
Thus, the number of the SHG susceptibilities is reduced from 8 to 6,
that is $\chi_{xxx}^{(2)}=-\chi_{xyy}^{(2)}=-\chi_{yyx}^{(2)}$, and
$\chi_{xxy}^{(2)}=\chi_{yxx}^{(2)}=-\chi_{yyy}^{(2)}$.
In order to compare these SHG susceptibilities with the results of the 2nd harmonic spectra, we focus on $\chi_{xxx}^{(2)}$.

Figure S3(a) shows the absolute value of $\chi_{xxx}^{(2)}$ for CrCl$_{3}$ with the damping parameter $\eta=0.0005$ Hartree.
It can be seen that the intensity of $|\chi_{xxx}^{(2)}|$ is nearly zero below 0.92 eV, which concides with the results of the 2nd harmonic spectra.
When the photon energy reaches 0.92 eV, the first resonance peak appears with an intensity of 1.05 pm/V.
The photon energy of the first peak is close to half of the energy gap of 1.69 eV for CrCl$_{3}$,
so the first peak can be regarded as an indicator of the band gap.
Compared to CrCl$_{3}$, the first peak of CrBr$_{3}$ is redshifted and appears at 0.75 eV, as shown in Fig. S3(b).
This is because the band gap of CrBr$_{3}$ is reduced to 1.42 eV.
Moreover, its intensity of the first peak is increased to 5.70 pm/V.
This means that the smaller the band gap, the larger the harmonic intensity.
This is also the case for CrI$_{3}$.
It has a smaller band gap of 0.84 eV and the intensity for the first peak is as large as 546.17 pm/V, as shown in Fig. S3(c).
Therefore, the first peaks in the cw SHG are consistent with the results obtained in the 2nd harmonic spectra.

\begin{figure}
\centering
\includegraphics[width=.6\textwidth]{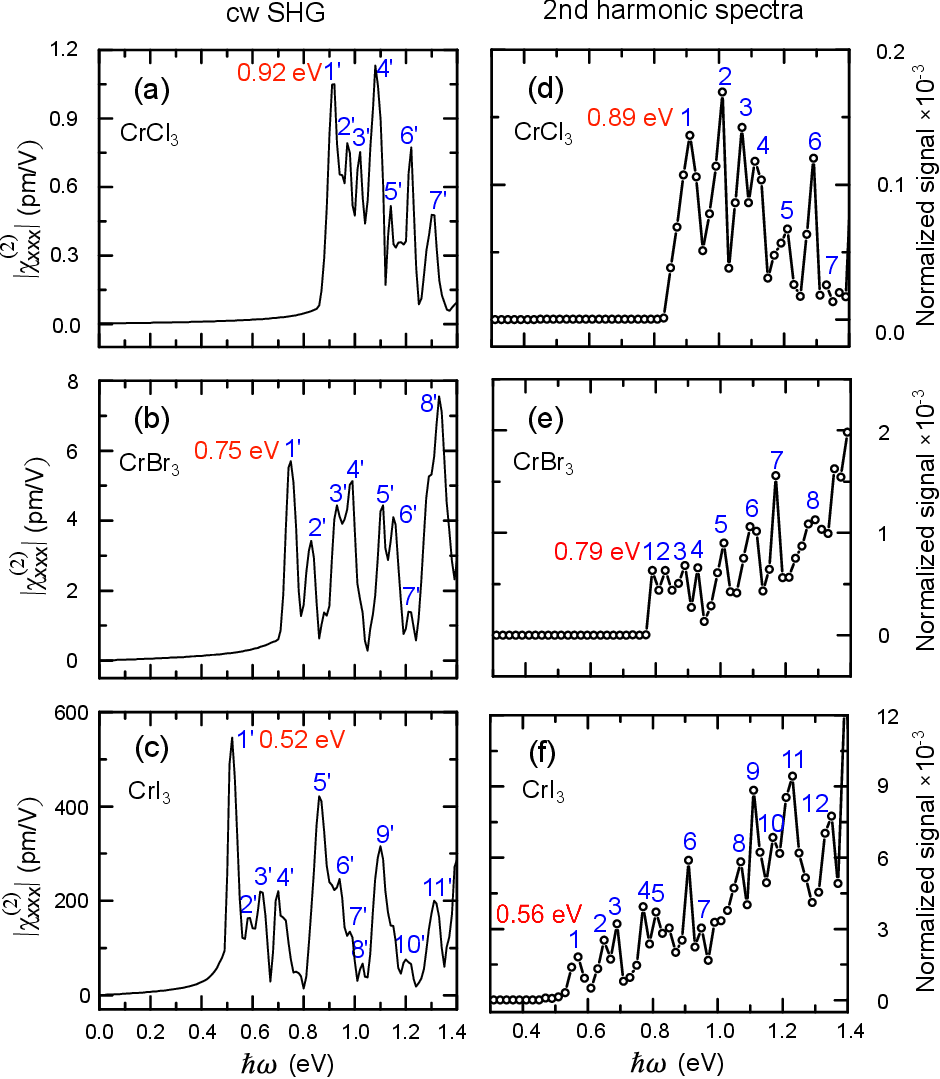}
\caption{The absolute value of $\chi_{xxx}^{(2)}$ in the energy range of 0$-$1.4 eV for bilayers (a) CrCl$_{3}$, (b) CrBr$_{3}$, and (c) CrI$_{3}$ with the damping parameter $\eta=0.0005$ Hartree.
The blue and red numbers near the peaks represent the order of the peaks and the photon energy of the first peak.
The 2nd harmonic spectra as a function of photon energy $\hbar\omega$ for
bilayers (d) CrCl$_{3}$, (e) CrBr$_{3}$, and (f) CrI$_{3}$.
The blue and red numbers near the peaks represent the order of the peaks and the photon energy of the first peak, respectively.
The laser polarization is along the $x$ axis, with the duration $\tau$ = 60 fs, and the vector potential amplitude $A_{0}$ = 0.03 Vfs/{\AA}.
}
\label{S2}
\end{figure}
When the photon energy is increased to 1.4 eV, many complex peaks appear.
For CrCl$_{3}$, there are six additional peaks 2$^{'}$$-$7$^{'}$ after the first peak, as shown in Fig. S3(a),
where we use prime on a number to denote the peaks calculated by the cw SHG susceptibility.
The photon energies of peaks 2$^{'}$ and 3$^{'}$ are 0.97 and 1.02 eV,
which are close to the photon energy 1.01 eV of peak 2 in the 2nd harmonic spectra, as shown in Fig. S3(d).
Peak 4$^{'}$ has the maximum intensity of 1.13 pm/V at 1.08 eV.
This indicates an enhanced electron transition at 2.16 eV.
This peak matches peak 3 with the photon energy of 1.07 eV in the 2nd harmonic spectra.
The photon energies of peaks 5$^{'}$$-$7$^{'}$ are 1.14, 1.22, and 1.31 eV,
which are close to the photon energies of 1.11, 1.21, and 1.29 eV for peaks 4, 5, and 6 in the 2nd harmonic spectra.
It is found that only the weak peak 7 in the 2nd harmonic spectra is not detected in the cw SHG.
We conclude that most of the peaks in the cw SHG are reproduced in the 2nd harmonic spectra for CrCl$_{3}$.
For the cw SHG of CrBr$_{3}$, there are seven peaks 2$^{'}$$-$8$^{'}$ after the first peak, as shown in Fig. S3(b).
The photon energies of peaks 2$^{'}$$-$6$^{'}$, and 8$^{'}$ are 0.83, 0.93, 0.99, 1.11, 1.15, and 1.33 eV,
matching peaks 2, and 4$-$8 in the 2nd harmonic spectra, as shown in Fig. S3(e).
The corresponding photon energies are 0.83, 0.93, 1.01, 1.09, 1.17, and 1.29 eV.
The intensity of peak 7$^{'}$ is the weakest and there is no corresponding peak in the 2nd harmonic spectra.
It is also found that peak 3 in the 2nd harmonic spectra has no matching peak in the cw SHG of CrBr$_{3}$.
In the case of CrI$_{3}$, the SHG spectrum has ten peaks 2$^{'}$$-$11$^{'}$ after the first peak, as shown in Fig. S3(c).
Seven peaks 3$^{'}$, 4$^{'}$, 6$^{'}$$-$9$^{'}$, and 11$^{'}$ are at 0.64, 0.7, 0.89, 0.94, 1.03, 1.1, and 1.32 eV,
which are close to peaks 2, 3, 6$-$9, and 12 in the 2nd harmonic spectra, as shown in Fig. S3(f).
The photon energies corresponding to these peaks are 0.65, 0.69, 0.91, 0.95, 1.07, 1.11, and 1.35 eV.
Peaks 1$^{'}$ and 2$^{'}$ with the photon energies of 0.52 and 0.59 eV match peak 1 in the 2nd harmonic spectra, and its photon energy is 0.57 eV.
The photon energy of peak 5$^{'}$ is 0.78 eV, which matches two peaks 4 and 5 in the 2nd harmonic spectra, and the corresponding photon energies are 0.77 and 0.81 eV.
Peak 10$^{'}$ is located at 1.2 eV, which is close to peaks 10 and 11 in the 2nd harmonic spectra at 1.17 and 1.23 eV.
This shows that the peaks in the cw SHG spectra are basically consistent with those in the 2nd harmonic spectra for these three materials,
indicating the reliability of the 2nd harmonic spectra to study the electronic transitions based on the band structure.

\begin{figure}
\centering
\includegraphics[width=.4\textwidth]{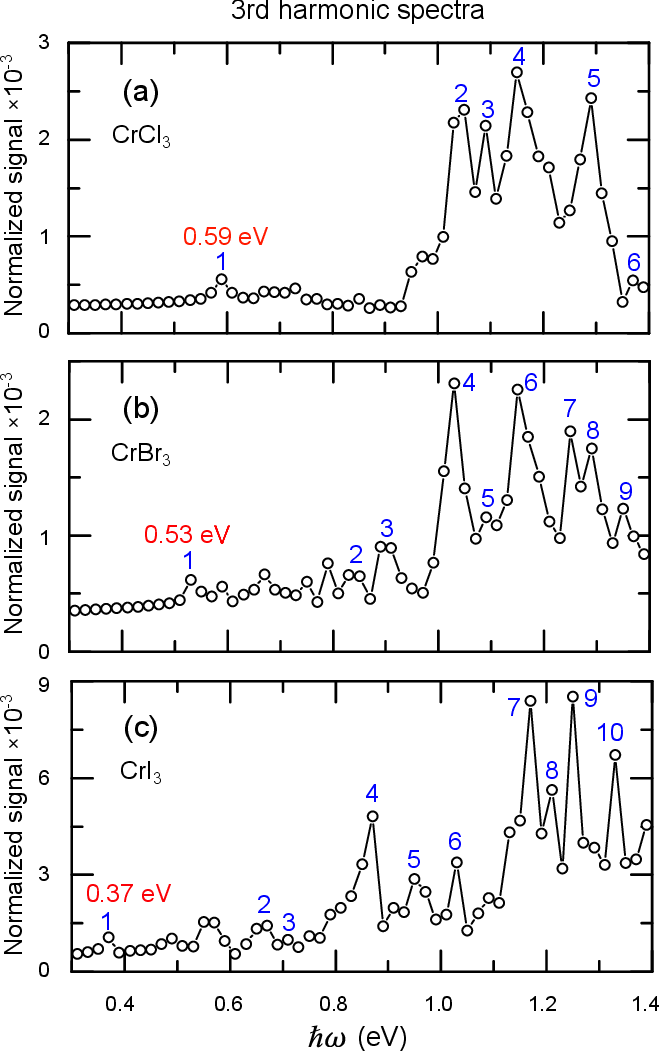}
\caption{
The 3rd harmonic spectra as a function of photon energy $\hbar\omega$ for
bilayers (a) CrCl$_{3}$, (b) CrBr$_{3}$, and (c) CrI$_{3}$.
The blue and red numbers near the peaks represent the order of the peaks and the photon energy of the first peak, respectively.
The laser polarization is along the $x$ axis, with the duration $\tau$ = 60 fs, and the vector potential amplitude $A_{0}$ = 0.03 Vfs/{\AA}.}
\label{S3}
\end{figure}

In addition, we also calculate the 3rd harmonic spectra for CrX$_{3}$.
Compared to the 2nd harmonic spectra, the 3rd harmonic spectra have a similar manner, as shown in Figs. S4(a)-S4(c).
For CrCl$_{3}$, the first peak appears at around 0.6 eV, owing to the real electron transition at 3$\hbar\omega$ of 1.8 eV.
This exactly reproduces the result of peak 1 in the 2nd harmonic spectra.
This also holds for CrBr$_{3}$ and CrI$_{3}$ with peaks 1 at 0.54 and 0.37 eV, respectively.
This implies that the 1st peak of 3rd harmonics spectra can be also as the indicator for the band gap of a material.
In the photon energy range of 0.6$-$0.9 eV, there also exist a series of peaks for CrCl$_{3}$, as shown in Fig. S4(a).
These peaks are consistent with those of 2nd harmonic spectra in the energy range of 0.9$-$1.4 eV.
Above 0.9 eV, we can see additional five peaks 2$-$6, which do not appear in the 2nd harmonics spectra.
These would be the eigenpeaks of 3rd harmonics spectra.
For CrBr$_{3}$ and CrI$_{3}$, the eigenpeaks are labeled as 2$-$9 and 2$-$10, respectively, as shown in Figs. S4(b) and S4(c).
Our calculated 3rd harmonic spectra would be a good guide for experimental researches.

\section{\uppercase\expandafter{\romannumeral4}. The underlying physics of eigenpeaks in the 2nd and 3rd harmonic spectra for CrBr$_{3}$}

\subsection{A. The 2nd harmonic spectra }

\begin{figure}[bh]
\centering
\includegraphics[width=.7\textwidth]{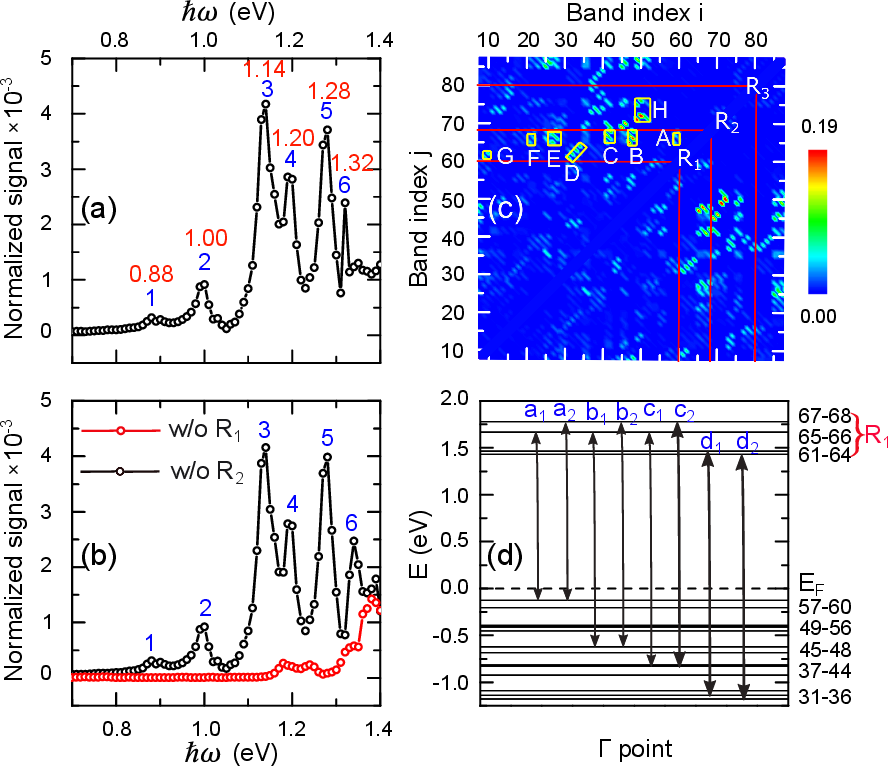}
\caption{(a) The 2nd harmonic spectra as a function of photon energy at $\Gamma$ for bilayer CrBr$_{3}$.
The blue and red numbers near each peak represent the peak indices and the corresponding photon energies, respectively.
(b) The 2nd harmonic spectra as a function of photon energy excluding $R_{1}$ (red line) and $R_{2}$ (black line) in the conduction bands.
(c) The absolute value of the momentum matrix elements $|\textbf{P}|$ for bands 1$-$88 in bilayer CrBr$_{3}$.
(d) The energy diagram at $\Gamma$ for bilayer CrBr$_{3}$,
where the arrows denote transitions with large momentum matrix elements,
and the letters on the arrow label the different transitions.
The band indices belonging to $R_{1}$ are marked on the right side of the figure.
The dashed line denotes the Fermi level. }
\label{S3}
\end{figure}

\renewcommand{\thetable}{S\uppercase\expandafter{\romannumeral1}}
\begin{table}[bh]
\centering
\caption{Transitions in CrBr$_{3}$ in regions A$-$D in Fig. S5(c) with large momentum matrix elements $|\textbf{P}|$.
The corresponding transition energies match the harmonic energies of the eigenpeaks,
where 1$-$6 denote those eigenpeaks in the 2nd harmonic spectra.
$\Delta$$E$ and 2$\hbar\omega$ represent the transition and harmonic energies, respectively, in unit of eV.
The momentum matrix elements are in unit of a.u..
}
\begin{tabular}{cccccccccccccccccccccccccc}\hline\hline
Region                    &&Transition                  &&$|\textbf{P}|$ &&$\Delta$$E$          && Peak                  && 2$\hbar\omega$ ($\hbar\omega$)                  \\ \hline
\multirow{2}{0.2cm}{A}  && 59,60$\rightarrow$65,66      &&0.06    &&\multirow{1}{1.5cm}{1.79 ($a_{1}$)} &&\multirow{1}{0.2cm}{1}    &&\multirow{1}{1.8cm}{1.76 (0.88)}                         \\
                        && 59,60$\rightarrow$67,68      &&0.07    &&\multirow{1}{1.5cm}{1.91 ($a_{2}$)} &&\multirow{1}{0.2cm}{2}    &&\multirow{1}{1.8cm}{2.00 (1.00)}                   \\\hline
\multirow{2}{0.2cm}{B}  && 47,48$\rightarrow$65,66      &&0.18    &&\multirow{1}{1.5cm}{2.28 ($b_{1}$)} &&\multirow{1}{0.2cm}{3}    &&\multirow{1}{1.8cm}{2.28 (1.14)}                   \\
                        && 47,48$\rightarrow$67,68      &&0.14    &&\multirow{1}{1.5cm}{2.40 ($b_{2}$)} &&\multirow{1}{0.2cm}{4}    &&\multirow{1}{1.8cm}{2.40 (1.20)}                       \\\hline
\multirow{2}{0.2cm}{C}  && 41,42$\rightarrow$65,66      &&0.12    &&\multirow{1}{1.5cm}{2.48 ($c_{1}$)} &&\multirow{1}{0.2cm}{4}    &&\multirow{1}{1.8cm}{2.40 (1.20)}                   \\
                        && 41,42$\rightarrow$67,68      &&0.15    &&\multirow{1}{1.5cm}{2.60 ($c_{2}$)} &&\multirow{1}{0.2cm}{5}    &&\multirow{1}{1.8cm}{2.56 (1.28)}                       \\\hline
\multirow{3}{0.2cm}{D}  && 33,34$\rightarrow$63,64      &&0.13    &&\multirow{1}{1.5cm}{2.60 ($d_{1}$)} &&\multirow{1}{0.2cm}{5}    &&\multirow{1}{1.8cm}{2.56 (1.28)}                   \\
                        &&\multirow{2}{2.1cm}{31,32$\rightarrow$61,62}&&\multirow{2}{0.7cm}{0.12}&&\multirow{2}{1.5cm}{2.61 ($d_{2}$)} &&\multirow{1}{0.2cm}{5}    &&\multirow{1}{1.8cm}{2.56 (1.28)}    \\
                        &&                 &&                 &&                                                                       &&\multirow{1}{0.2cm}{6}    &&\multirow{1}{1.8cm}{2.64 (1.32)}     \\
\hline\hline
\end{tabular}\label{tabS1}
\end{table}

The eigenpeaks in the 2nd harmonic spectra are intrinsic of
a material, which largely depend on the band structure and the momentum matrix elements $\textbf{P}$.
To unveil the relation between them, we calculate the 2nd harmonic spectra of CrBr$_{3}$ at $\Gamma$, as shown in Fig. S5(a).
We scan the incident photon energies between 0.7 and 1.4 eV in step of 0.01 eV.
When the photon energy is below 0.88 eV, the harmonic amplitudes are close to zero,
indicating that the electron transitions from the valence bands to the conduction bands are virtual.
When the photon energy reaches 0.88 eV, the first peak appears.
This means that real transitions touch the bands, and the energy difference between bands is close to $2\hbar\omega=1.76$ eV.
With the further increase of photon energy, there are five peaks 2$-$6.
The amplitude of peak 2 is slightly larger than that of peak 1.
Peaks 3$-$6 are much larger, and their photon energies are 1.14, 1.20, 1.28, and 1.32 eV, respectively.
This means that the enhanced electron transitions would occur in the energy differences of 2.28, 2.40, 2.56, and 2.64 eV.

As mentioned above, the conduction bands for CrBr$_{3}$ have three individual regions $R_{1}$$-$$R_{3}$.
Similarly, we study the effect of these three regions $R_{1}$$-$$R_{3}$ on those eigenpeaks of the 2nd harmonic spectra.
Excluding $R_{1}$, the 2nd harmonic spectra are shown in the red line of Fig. S5(b).
It is found that peaks 1$-$6 decrease significantly.
This means that the transitions between the valence bands and the conduction bands in $R_{1}$ dominate peaks 1$-$6.
When $R_{2}$ is excluded, peaks 1$-$6 almost remain unchanged, as shown in the black line of Fig. S5(b).
This confirms peaks 1$-$6 are related to the transitions between the valence bands and the conduction bands in $R_{1}$.
Only $R_{3}$ is excluded, all peaks remain unchanged, indicating that $R_{3}$ has no effect on peaks 1$-$6.

To further explain the underlying physics, we need to resort to the momentum matrix elements $|\textbf{P}|$ and the band structures, as shown in Figs. S5(c) and S5(d).
In Fig. S5(c), there are four regions A$-$D with large momentum matrix elements between the valence bands and the conduction bands in $R_{1}$.
Region A has two transitions $a_{1}$ and $a_{2}$, as shown in Fig. S5(d) and Table SI.
The transition $a_{1}$ is between the valence bands 59 and 60 and the conduction bands 65 and 66,
where the valence and conduction bands are degenerate, marked as 59,60$\rightarrow$65,66 in Table S\uppercase\expandafter{\romannumeral1}.
The transition energies of $a_{1}$ and $a_{2}$ are 1.79 and 1.91 eV, which are close to the harmonic energies of 1.76 and 2.00 eV for peaks 1 and 2, respectively.
The momentum matrix elements $|\textbf{P}|$ of these two peaks are only 0.06 and 0.07 a.u., respectively, as shown in Table SI.
This confirms that the amplitudes of peaks 1 and 2 are small.
Two transitions $b_{1}$ and $b_{2}$ in region B have transition energies of 2.28 and 2.40 eV, respectively.
The transition energy of $b_{1}$ matches the harmonic energy of 2.28 eV for peak 3, indicating that peak 3 is dominated by $b_{1}$.
The transition energy of $b_{2}$ is equal to the harmonic energy of 2.40 eV for peak 4.
This means that peak 4 is mainly contributed by $b_{2}$.
Region C contains two transitions $c_{1}$ and $c_{2}$.
The transition energy of $c_{1}$ is 2.48 eV, which is very close to the harmonic energy of peak 4.
The transition energy of $c_{2}$ is 2.60 eV, and the harmonic energy of peak 5 is 2.56 eV.
Therefore, peak 5 is mainly contributed by $c_{2}$.
For region D, there are two transitions $d_{1}$ and $d_{2}$ with the transition energies of 2.60 and 2.61 eV, respectively.
Their transition energies match peaks 5 and 6 with the harmonic energies of 2.56 and 2.64 eV, respectively.
The amplitudes of peaks 3$-$6 are relatively strong, which originates from the large $|\textbf{P}|$ of the corresponding transitions, as shown in Table SI.

\subsection{B. The 3rd harmonic spectra }

\begin{figure}[bh]
\centering
\includegraphics[width=.7\textwidth]{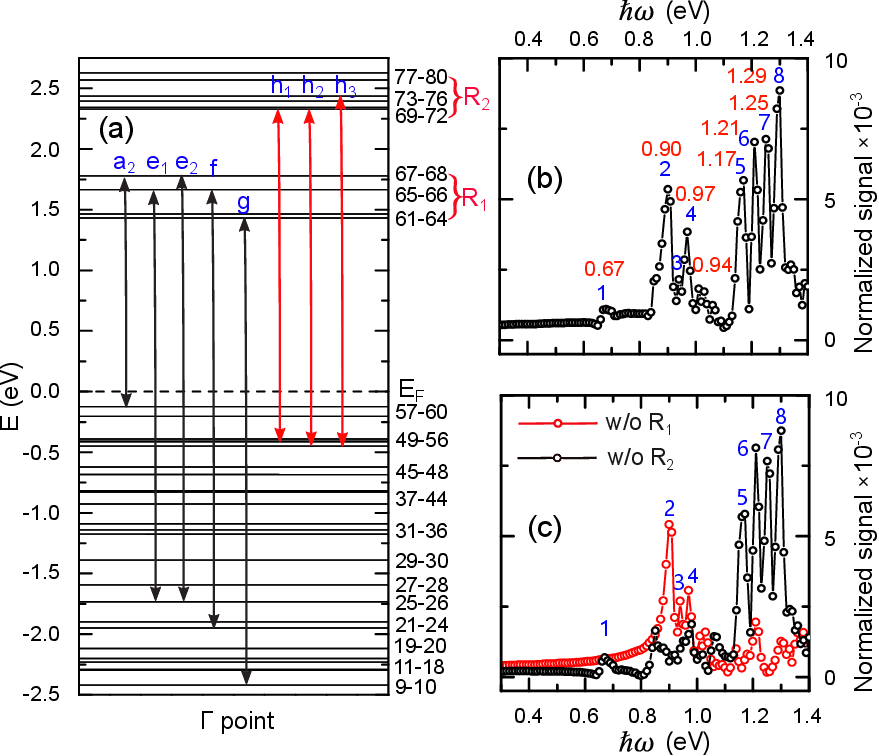}
\caption{(a) The energy diagram at $\Gamma$ for bilayer CrBr$_{3}$,
where the arrows denote transitions with large momentum matrix elements,
and the letters on the arrow label different transitions.
The band indices belonging to $R_{1}$ and $R_{2}$ are marked on the right side of the figure.
The dashed line denotes the Fermi level.
(b) The 3rd harmonic spectra as a function of photon energy $\hbar\omega$ at $\Gamma$.
The blue and red numbers near each peak represent the peak indices and the corresponding photon energies, respectively.
(c) The 3rd harmonic spectra as a function of photon energy excluding $R_{1}$ (red line) and $R_{2}$ (black line) in the conduction bands.
 }
\label{S4}
\end{figure}

For bilayer CrBr$_{3}$, the eigenpeaks of the 3rd harmonic spectra come from the resonant transitions with a large energy difference
between the valence and conduction bands, as shown in Fig. S6(a).
The 3rd harmonic spectra at $\Gamma$ are displayed in Fig. S6(b),
and the incident photon energies are between 0.3 and 1.4 eV in step of 0.01 eV.
It is found that the photon energy of peak 1 is 0.67 eV, which is close to half of the band gap 1.42 eV.
Peaks 2$-$4 are close to each other, and the amplitude of peak 2 is stronger than that of the other two peaks.
The amplitudes of peaks 5$-$8 are relatively large, and their photon energies are 1.17, 1.21, 1.25, and 1.29 eV, respectively.
Next, we determine the effect of the transitions between the valence bands and the conduction bands in $R_{1}$$-$$R_{3}$ on these eigenpeaks.
When $R_{1}$ is excluded, the 3rd harmonic spectra are displayed in the red line of Fig. S6(c).
It is shown that peaks 1, 5$-$8 are significantly reduced, and peaks 2$-$4 have nearly changed.
This means that the reduced peaks 1, 5$-$8 are mainly contributed by the transitions between the valence bands and the conduction bands in $R_{1}$.
Excluding $R_{2}$, the corresponding harmonic spectra are shown in the black line of Fig. S6(c).
It is found that peaks 2$-$4 decrease, while peaks 1, 5$-$8 keep unchanged.
This indicates that peaks 2$-$4 are mainly contributed by the transitions between the valence bands and the conduction bands in $R_{2}$.
When $R_{3}$ is excluded, and $R_{1}$ and $R_{2}$ are still retained, all peaks have almost unchanged,
indicating that $R_{3}$ has no effect on these eigenpeaks.

\renewcommand{\thetable}{S\uppercase\expandafter{\romannumeral2}}
\begin{table}[bt]
\centering
\caption{Transitions in CrBr$_{3}$ in regions A, E$-$H in Fig. S5(c) with large momentum matrix elements $|\textbf{P}|$.
The corresponding transition energies match the harmonic energies of the eigenpeaks,
where 1$-$8 denote those eigenpeaks in the 3rd harmonic spectra.
$\Delta$$E$ and 3$\hbar\omega$ represent the transition and harmonic energies, respectively, in unit of eV.
The momentum matrix elements are in unit of a.u..
}
\begin{tabular}{cccccccccccccccccccccccccc}\hline\hline
Region                    &&Transition                  &&$|\textbf{P}|$ &&$\Delta$$E$          && Peak                  && 3$\hbar\omega$ ($\hbar\omega$)                                   \\ \hline
\multirow{1}{0.2cm}{A}  && 59,60$\rightarrow$67,68      &&0.07    &&\multirow{1}{1.5cm}{1.91 ($a_{2}$)} &&\multirow{1}{0.2cm}{1}    &&\multirow{1}{1.8cm}{2.01 (0.67)}                        \\\hline
\multirow{2}{0.2cm}{E}  && 25,26$\rightarrow$66,65      &&0.10    &&\multirow{1}{1.5cm}{3.41 ($e_{1}$)} &&\multirow{1}{0.2cm}{5}    &&\multirow{2}{1.8cm}{3.51 (1.17)}                         \\
                        && 25,26$\rightarrow$68,67      &&0.10    &&\multirow{1}{1.5cm}{3.51 ($e_{2}$)} &&\multirow{1}{0.2cm}{5}    &&                                                          \\\hline
\multirow{1}{0.2cm}{F}  && 21,22$\rightarrow$65,66      &&0.11    &&\multirow{1}{1.5cm}{3.61 ($f$)}       &&\multirow{1}{0.2cm}{6}    &&\multirow{1}{1.8cm}{3.63 (1.21)}                         \\\hline
\multirow{2}{0.2cm}{G}  &&\multirow{2}{1.8cm}{9,10$\rightarrow$61,62}&&\multirow{2}{0.7cm}{0.11}&&\multirow{2}{1.5cm}{3.85 (g)}  &&\multirow{1}{0.2cm}{7}    &&\multirow{1}{1.8cm}{3.75 (1.25)}   \\
                        &&            &&        &&                                                                               &&\multirow{1}{0.2cm}{8}    &&\multirow{1}{1.8cm}{3.87 (1.29)}       \\\hline
\multirow{3}{0.2cm}{H}  &&\multirow{1}{2.1cm}{49,50$\rightarrow$71,72}&&\multirow{1}{0.7cm}{0.21}&&\multirow{1}{1.5cm}{2.75 ($h_{1}$)} &&\multirow{1}{0.2cm}{2}    &&\multirow{1}{1.8cm}{2.70 (0.90)}  \\
                        &&\multirow{1}{2.1cm}{51,52$\rightarrow$72,71}&&\multirow{1}{0.7cm}{0.12}&&\multirow{1}{1.5cm}{2.79 ($h_{2}$)} &&\multirow{1}{0.2cm}{3}    &&\multirow{1}{1.8cm}{2.82 (0.94)}    \\
                        &&\multirow{1}{2.1cm}{49,50$\rightarrow$75,76}&&\multirow{1}{0.7cm}{0.13}&&\multirow{1}{1.5cm}{2.89 ($h_{3}$)} &&\multirow{1}{0.2cm}{4}    &&\multirow{1}{1.8cm}{2.91 (0.97)}      \\
\hline\hline
\end{tabular}\label{tab1}
\end{table}

Similar to the 2nd harmonic spectra, those eigenpeaks in the 3rd harmonic spectra are also closely related to the large momentum matrix elements $\textbf{P}$.
The corresponding $\textbf{P}$ between the valence bands and the conduction bands in $R_{1}$ has five regions, labeled A, E$-$H, as shown in Fig. S5(c).
The transition $a_{2}$ is one of the transitions in region A as shown in Fig. S6(a) and Table S\uppercase\expandafter{\romannumeral2}.
The transition energy of the transition $a_{2}$ is 1.91 eV,
which is close to the harmonic energy of peak 1 with $3\hbar\omega=2.01$ eV.
The absolute value of the momentum matrix element $|\textbf{P}|$ is 0.07 a.u..
This can explain the small amplitude of peak 1.
Region E contains two transitions $e_{1}$ and $e_{2}$ with the transition energies of 3.41 and 3.51 eV, respectively.
These two transition energies almost match the harmonic energy of peak 5 with $3\hbar\omega=3.51$ eV.
In region F, the transition energy of the transition $f$ is 3.61 eV, involving peak 6 with the harmonic energy of 3.63 eV.
The transition energy of the transition $g$ in region G is 3.85 eV, which matches peaks 7 and 8 with the harmonic energies of 3.75 and 3.87 eV, respectively.
The momentum matrix elements $|\textbf{P}|$ of transitions $e_{1}$, $e_{2}$, $f$, and $g$ are 0.10, 0.10, 0.11, and 0.11 eV, respectively, which are comparable.
This leads to an equivalent amplitude of the corresponding peaks.
There is only one region H related to the transitions between the valence bands and the conduction bands in $R_{2}$, as shown in Fig. S5(c).
In region H, the transition energies of three transitions $h_{1}$, $h_{2}$, and $h_{3}$ are 2.75, 2.79, and 2.89 eV, respectively.
These three values are close to the harmonic energies of 2.70, 2.82, and 2.91 eV for peaks 2, 3, and 4,
indicating that these three peaks are dominated by the transitions $h_{1}$, $h_{2}$, and $h_{3}$.
The momentum matrix elements $|\textbf{P}|$ of the transitions $h_{1}$, $h_{2}$, and $h_{3}$ are 0.21, 0.12, and 0.13 a.u., as shown in Table S\uppercase\expandafter{\romannumeral2}.
It is noted that the momentum matrix elements $|\textbf{P}|$ of the transition $h_{1}$ corresponding to peak 2 is the largest among these three values,
while the momentum matrix elements $|\textbf{P}|$ of the transitions $h_{2}$ and $h_{3}$ corresponding to peaks 3 and 4 are close to each other.
This indicates that the amplitude of peak 2 is larger than that of peaks 3 and 4, and the amplitudes of peaks 3 and 4 are comparable.

\section{\uppercase\expandafter{\romannumeral5}. The underlying physics of eigenpeaks in the 2nd and 3rd harmonic spectra for CrI$_{3}$}

\subsection{A. The 2nd harmonic spectra}

\begin{figure}[bh]
\centering
\includegraphics[width=.7\textwidth]{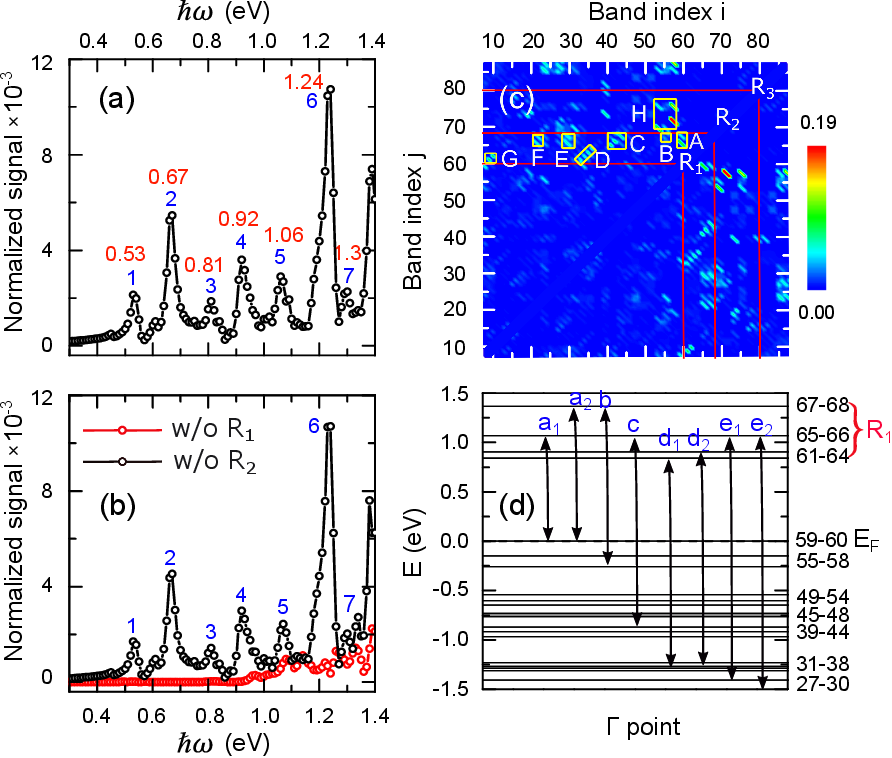}
\caption{(a) The 2nd harmonic spectra as a function of photon energy at $\Gamma$ for bilayer CrI$_{3}$.
The blue and red numbers near each peak represent the peak indices and the corresponding photon energies, respectively.
(b) The 2nd harmonic spectra as a function of photon energy excluding $R_{1}$ (red line) and $R_{2}$ (black line) in the conduction bands.
(c) The absolute value of the momentum matrix elements $|\textbf{P}|$ for bands 1$-$88 in bilayer CrI$_{3}$.
(d) The energy diagram at $\Gamma$ for bilayer CrI$_{3}$,
where the arrows denote transitions with large momentum matrix elements,
and the letters on the arrow label the different transitions.
The band indices belonging to $R_{1}$ are marked on the right side of the figure.
The dashed line denotes the Fermi level. }
\label{S6}
\end{figure}

We also calculate the 2nd harmonic spectra of CrI$_{3}$, as shown in Fig. S7(a).
We scan the incident photon energies between 0.3 and 1.4 eV in step of 0.01 eV.
It is shown that there are seven peaks 1$-$7. The first peak appears at 0.53 eV.
The corresponding harmonic energy is about 0.22 eV larger than the band gap.
The photon energy of the strong peak 6 is 1.25 eV,
indicating that the enhanced electron transitions would appear in the energy difference of 2.50 eV.
The small peaks 2$-$5 and 7 are at 0.67, 0.81, 0.92, 1.06 and 1.30 eV.
Similarly, we mainly focus on the effect of the transitions between the valence bands and the conduction bands in $R_{1}$$-$$R_{3}$ on the eigenpeaks.
When $R_{1}$ is excluded, the 2nd harmonic spectra are shown in the red line of Fig. S7(b).
It is found that peaks 1$-$3 completely disappear and peaks 4$-$7 decrease significantly.
Peaks 1$-$7 are mainly contributed by the transitions between the valence bands and the conduction bands in $R_{1}$.
To confirm this result, $R_{2}$ is excluded, and all peaks remain unchanged, as shown in the black line of Fig. S7(b).
When only $R_{3}$ is excluded, all peaks 1$-$7 remain unchanged.
This indicates that peaks 1$-$7 are only related to the transitions between the valence bands and the conduction bands in $R_{1}$.

\renewcommand{\thetable}{S\uppercase\expandafter{\romannumeral3}}
\begin{table}[bt]
\centering
\caption{Transitions in CrI$_{3}$ in regions A$-$E in Fig. S7(c) with large momentum matrix elements $|\textbf{P}|$.
The corresponding transition energies match the harmonic energies of the eigenpeaks,
where 1$-$7 denote those eigenpeaks in the 2nd harmonic spectra.
$\Delta$$E$ and 2$\hbar\omega$ represent the transition and harmonic energies, respectively, in unit of eV.
The momentum matrix elements are in unit of a.u..
}
\begin{tabular}{cccccccccccccccccccccccccc}\hline\hline
Region                    &&Transition  &&$|\textbf{P}|$ &&$\Delta$$E$                                    && Peak                     && 2$\hbar\omega$ ($\hbar\omega$)     \\ \hline
\multirow{2}{0.2cm}{A}  && 59,60$\rightarrow$66,65      &&0.07    &&\multirow{1}{1.5cm}{1.07 ($a_{1}$)} &&\multirow{1}{0.2cm}{1}    &&\multirow{1}{1.8cm}{1.06 (0.53)}   \\
                        && 59,60$\rightarrow$68,67      &&0.11    &&\multirow{1}{1.5cm}{1.36 ($a_{2}$)} &&\multirow{1}{0.2cm}{2}    &&\multirow{1}{1.8cm}{1.34 (0.67)}    \\\hline
\multirow{1}{0.2cm}{B}  && 55,56$\rightarrow$67,68      &&0.07    &&\multirow{1}{1.5cm}{1.62 ($b$)}     &&\multirow{1}{0.2cm}{3}    &&\multirow{1}{1.8cm}{1.62 (0.81)}     \\\hline
\multirow{1}{0.2cm}{C}  && 43,44$\rightarrow$66,65      &&0.07    &&\multirow{1}{1.5cm}{1.93 ($c$)}     &&\multirow{1}{0.2cm}{4}    &&\multirow{1}{1.8cm}{1.84 (0.92)}      \\\hline
\multirow{2}{0.2cm}{D}  && 33,34$\rightarrow$61,62      &&0.10    &&\multirow{1}{1.5cm}{2.12 ($d_{1}$)} &&\multirow{1}{0.2cm}{5}    &&\multirow{1}{1.8cm}{2.12 (1.06)}       \\
                        && 35,36$\rightarrow$63,64      &&0.11    &&\multirow{1}{1.5cm}{2.17 ($d_{2}$)} &&\multirow{1}{0.2cm}{5}    &&\multirow{1}{1.8cm}{2.12 (1.06)}        \\\hline
\multirow{2}{0.2cm}{E}  && 29,30$\rightarrow$66,65      &&0.15    &&\multirow{1}{1.5cm}{2.48 ($e_{1}$)} &&\multirow{1}{0.2cm}{6}    &&\multirow{1}{1.8cm}{2.50 (1.25)}    \\
                        && 27,28$\rightarrow$66,65      &&0.07    &&\multirow{1}{1.5cm}{2.61 ($e_{2}$)} &&\multirow{1}{0.2cm}{7}    &&\multirow{1}{1.8cm}{2.60 (1.30)}           \\
\hline\hline
\end{tabular}\label{tab1}
\end{table}

These eigenpeaks are closely related to the large momentum matrix elements $\textbf{P}$.
Large $\textbf{P}$ related to $R_{1}$ has five regions, labeled A, B, C, D, and E, as shown in Fig. S7(c).
Region A contains two transitions $a_{1}$ and $a_{2}$, as shown in Fig. S7(d) and Table S\uppercase\expandafter{\romannumeral3}.
Their transition energies are 1.07 and 1.36 eV, respectively.
These are close to the harmonic energies of peaks 1 and 2, which are 1.06 and 1.34 eV, respectively.
The momentum matrix elements $|\textbf{P}|$ for $a_{1}$ and $a_{2}$ are 0.07 and 0.11 a.u., respectively.
The transition energy of $b$ in region B is 1.62 eV, which exactly matches the harmonic energy of peak 3.
In region C, there is a transition $c$ with the transition energy of 1.93 eV.
The harmonic energy of peak 4 is 1.84 eV, which is close to the transition energy of $c$, so peak 4 is contributed by the transition $c$.
Region D has two transitions $d_{1}$ and $d_{2}$, and their transition energies are 2.12 and 2.17 eV, respectively.
These values match the harmonic energy of 2.12 eV for peak 5.
The transitions $b$, $c$, $d_{1}$, and $d_{2}$ have smaller $|\textbf{P}|$ of 0.07, 0.07, 0.10, and 0.11 a.u..
Region E contains two transitions $e_{1}$ and $e_{2}$ with transition energies of 2.48 and 2.61 eV, respectively, matching peaks 6 and 7.
The harmonic energies of peaks 6 and 7 are 2.50 and 2.60 eV, respectively.
This means that peaks 6 and 7 are contributed by the transitions $e_{1}$ and $e_{2}$, respectively.
$|\textbf{P}|$ for these two transitions $e_{1}$ and $e_{2}$ are 0.15 and 0.07 a.u..
$|\textbf{P}|$ of peak 6 is larger than that of peak 7, which is consistent with their peak amplitudes.

\subsection{B. The 3rd harmonic spectra }

Next, we focus on the 3rd harmonic spectra of CrI$_{3}$.
The energy difference between the valence and conduction bands corresponding to the eigenpeaks is shown in Fig. S8(a).
Figure S8(b) shows the 3rd harmonic spectra at $\Gamma$,
where the incident photon energies are between 0.3 and 1.4 eV in step of 0.01 eV.
It is found that the photon energy of the first peak is 0.35 eV, and the corresponding harmonic energy is close to the band gap of 0.84 eV.
As the photon energy gradually increases to 0.69 eV, there are still seven peaks 2$-$8.
The positions of peaks 2 and 3, peaks 4 and 5, peaks 6, 7, and 8 are close to each other, respectively.
Among them, the amplitude of peak 5 is the largest one.

\begin{figure}
\centering
\includegraphics[width=.7\textwidth]{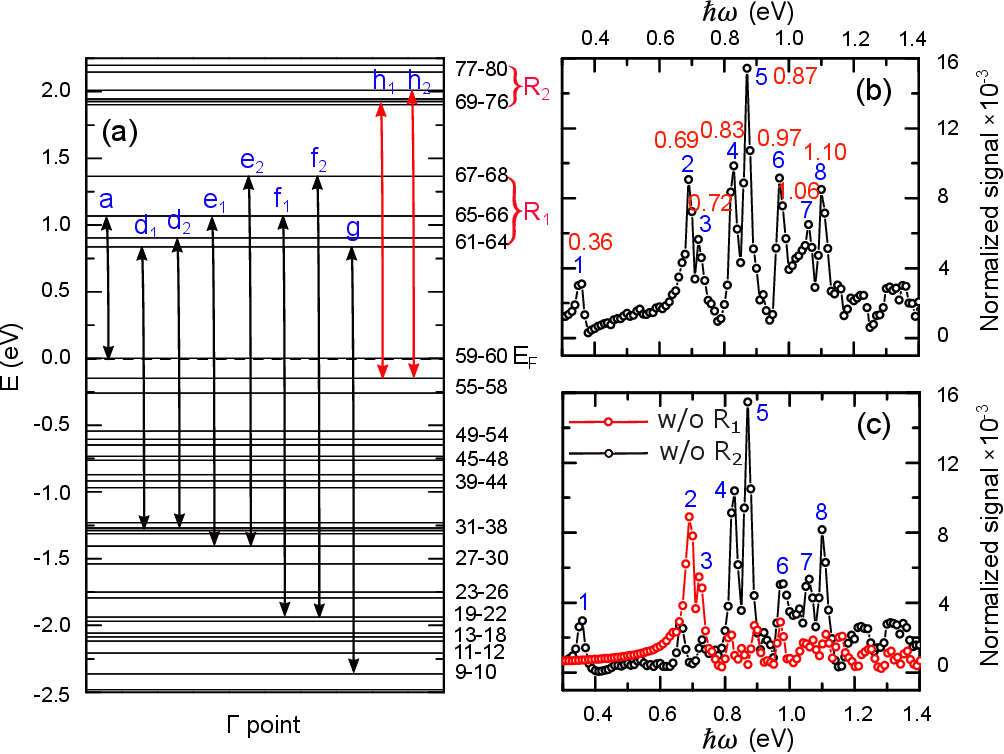}
\caption{(a) The energy diagram at $\Gamma$ for bilayer CrI$_{3}$,
where the arrows denote transitions with large momentum matrix elements,
and the letters on the arrow label different transitions.
The band indices belonging to $R_{1}$ and $R_{2}$ are marked on the right side of the figure.
The dashed line denotes the Fermi level.
(b) The 3rd harmonic spectra as a function of photon energy $\hbar\omega$ at $\Gamma$.
The blue and red numbers near each peak represent the peak indices and the corresponding photon energies, respectively.
(c) The 3rd harmonic spectra as a function of photon energy excluding $R_{1}$ (red line) and $R_{2}$ (black line) in the conduction bands.
 }
\label{S7}
\end{figure}
The conduction bands for three regions $R_{1}$$-$$R_{3}$ have an effect on these eigenpeaks.
When $R_{1}$ is excluded, six peaks 1, 4$-$8 decrease significantly,
while peaks 2 and 3 slightly reduced, as shown in the red line of Fig. S8(c).
This means that the transitions between the valence bands and the conduction bands in $R_{1}$ dominant peaks 1, 4$-$8.
The transitions between the valence bands and those in $R_{1}$ contribute little to peaks 2 and 3.
Excluding $R_{2}$, peaks 2 and 3 obviously decrease,
the additional six peaks are nearly unchanged, as shown in the black line of Fig. S8(c).
It can be obtained that the peaks 2 and 3 are mainly contributed by the transitions between the valence bands and the conduction bands in $R_{2}$.
When $R_{1}$ and $R_{2}$ are retained and $R_{3}$ is excluded, peaks 1$-$8 are almost unchanged,
indicating that these peaks are not related to $R_{3}$.

\renewcommand{\thetable}{S\uppercase\expandafter{\romannumeral4}}
\begin{table}
\centering
\caption{Transitions in CrI$_{3}$ in regions A, D$-$H in Fig. S7(c) with large momentum matrix elements $|\textbf{P}|$.
The corresponding transition energies match the harmonic energies of the eigenpeaks,
where 1$-$8 denote those eigenpeaks in the 3rd harmonic spectra.
$\Delta$$E$ and 3$\hbar\omega$ represent the transition and harmonic energies, respectively, in unit of eV.
The momentum matrix elements are in unit of a.u..
}
\begin{tabular}{cccccccccccccccccccccccccc}\hline\hline
Region                    &&Transition  &&$|\textbf{P}|$ &&$\Delta$$E$                                    && Peak                     && 3$\hbar\omega$ ($\hbar\omega$)                 \\ \hline
\multirow{1}{0.2cm}{A}  && 59,60$\rightarrow$66,65      &&0.07    &&\multirow{1}{1.5cm}{1.07 ($a_{1}$)} &&\multirow{1}{0.2cm}{1}    &&\multirow{1}{1.8cm}{1.08 (0.36)}          \\\hline
\multirow{2}{0.2cm}{D}  && 33,34$\rightarrow$61,62      &&0.10    &&\multirow{1}{1.5cm}{2.12 ($d_{1}$)} &&\multirow{1}{0.2cm}{2}    &&\multirow{1}{1.8cm}{2.07 (0.69)}       \\
                        && 35,36$\rightarrow$63,64      &&0.10    &&\multirow{1}{1.5cm}{2.17 ($d_{2}$)} &&\multirow{1}{0.2cm}{3}    &&\multirow{1}{1.8cm}{2.16 (0.72)}       \\\hline
\multirow{2}{0.2cm}{E}  && 29,30$\rightarrow$66,65      &&0.10    &&\multirow{1}{1.5cm}{2.48 ($e_{1}$)} &&\multirow{1}{0.2cm}{4}    &&\multirow{1}{1.8cm}{2.49 (0.83)}         \\
                        && 27,28$\rightarrow$66,65      &&0.13    &&\multirow{1}{1.5cm}{2.61 ($e_{2}$)} &&\multirow{1}{0.2cm}{5}    &&\multirow{1}{1.8cm}{2.61 (0.87)}          \\\hline
\multirow{2}{0.2cm}{F}  && 23,24$\rightarrow$66,65      &&0.10    &&\multirow{1}{1.5cm}{2.86 ($f_{1}$)} &&\multirow{1}{0.2cm}{6}    &&\multirow{1}{1.8cm}{2.91 (0.97)}\\
                        && 21,22$\rightarrow$68,67      &&0.10    &&\multirow{1}{1.5cm}{3.30 ($f_{2}$)} &&\multirow{1}{0.2cm}{8}    &&\multirow{1}{1.8cm}{3.30 (1.10)}          \\\hline
\multirow{1}{0.2cm}{G}  && 9,10$\rightarrow$61,62       &&0.10    &&\multirow{1}{1.5cm}{3.20 ($g$)}     &&\multirow{1}{0.2cm}{7}    &&\multirow{1}{1.8cm}{3.18 (1.06)}\\\hline
\multirow{2}{0.2cm}{H}  && 57,58$\rightarrow$72,71      &&0.23    &&\multirow{1}{1.5cm}{2.07 ($h_{1}$)} &&\multirow{1}{0.2cm}{2}    &&\multirow{1}{1.8cm}{2.07 (0.69)}      \\
                        && 57,58$\rightarrow$76,75      &&0.16    &&\multirow{1}{1.5cm}{2.15 ($h_{2}$)} &&\multirow{1}{0.2cm}{3}    &&\multirow{1}{1.8cm}{2.16 (0.72)}       \\
\hline\hline
\end{tabular}\label{tab1}
\end{table}

The underlying physics is related to the momentum matrix elements $|\textbf{P}|$.
The large $|\textbf{P}|$ related to $R_{1}$ have six regions A, D$-$G.
In region A, the transition $a_{1}$ is shown in Fig. S8(a) and Table S\uppercase\expandafter{\romannumeral4}. Its transition energy is 1.07 eV.
This transition energy is close to the harmonic energy of 1.08 eV for peak 1, indicating that peak 1 is dominated by the transition $a_{1}$.
The corresponding momentum matrix element is 0.07 a.u..
In region D, there are two transitions $d_{1}$ and $d_{2}$, and their transition energies are 2.12 and 2.17 eV, respectively.
They match peaks 2 and 3 with the harmonic energies of 2.07 and 2.16 eV.
The $|\textbf{P}|$ of transitions $d_{1}$ and $d_{2}$ is equal to 0.10 a.u..
The transitions $e_{1}$ and $e_{2}$ are included in region E.
The transition energy of $e_{1}$ is 2.48 eV, which is close to peak 4 with $3\hbar\omega=2.49$ eV.
The transition energy of $e_{2}$ is 2.61 eV, which matches peak 5 with $3\hbar\omega=2.61$ eV.
For region F, the transition energies of $f_{1}$ and $f_{2}$ are 2.86 eV and 3.30 eV, respectively.
The harmonic energies of peaks 6 and 8 are 2.91 and 3.30 eV, respectively, which match transitions $f_{1}$ and $f_{2}$.
It is obtained that peaks 6 and 8 are dominated by $f_{1}$ and $f_{2}$, respectively.
There is only one transition $g$ in region G, and its transition energy is 3.20 eV.
It matches the harmonic energy of peak 7 with $3\hbar\omega=3.18$ eV.
The momentum matrix elements $|\textbf{P}|$ of transitions $f_{1}$, $f_{2}$, and $g$ are 0.10, 0.10, and 0.10 a.u., respectively.
This means that the amplitudes of three peaks 6, 7, and 8 are comparable.

For the large momentum matrix elements $|\textbf{P}|$ related to $R_{2}$,
region H contains two transitions $h_{1}$ and $h_{2}$, as shown in Fig. S8(a) and Table S\uppercase\expandafter{\romannumeral4}.
The transition energy of $h_{1}$ is 2.07 eV, which matches peak 2 with $3\hbar\omega=2.07$ eV.
The harmonic energy of peak 3 is 2.16 e V, which is contributed by the transition $h_{2}$, and its transition energy is 2.15 eV.
This means that peaks 2 and 3 are dominated by the transitions $h_{1}$ and $h_{2}$.
The momentum matrix elements $|\textbf{P}|$ of $h_{1}$ and $h_{2}$ are 0.23 and 0.16 a.u..
which are almost twice as large as those of $d_{1}$ and $d_{2}$.
This confirms that peaks 2 and 3 are mainly contributed by the transitions between the valence bands to the conduction bands in $R_{2}$,
while the transitions between the valence bands and those bands in $R_{1}$ contribute little to these two peaks.